\documentclass{article}

\usepackage{microtype}
\usepackage{multirow}
\usepackage{graphicx}
\usepackage{subfigure}
\usepackage{booktabs}
\usepackage{subcaption}
\usepackage[export]{adjustbox}
\usepackage{hyperref}

\usepackage[preprint]{icml2025}
\usepackage{amsmath}
\usepackage{amssymb}
\usepackage{mathtools}
\usepackage{amsthm}

\usepackage[capitalize,noabbrev]{cleveref}

\theoremstyle{plain}

\theoremstyle{definition}

\theoremstyle{remark}

\usepackage[textsize=tiny]{todonotes}

\definecolor{darkblue}{rgb}{0.0, 0.0, 0.55}
\icmltitlerunning{Conformation-Aware Structure Prediction of Antigen-Recognizing Immune Proteins}

\begin{document}
\twocolumn[
\icmltitle{Conformation-Aware Structure Prediction of Antigen-Recognizing\\ Immune Proteins}

\begin{icmlauthorlist}
\icmlauthor{Fr\'ed\'eric A. Dreyer}{pd}
\icmlauthor{Jan Ludwiczak}{pd}
\icmlauthor{Karolis Martinkus}{pd}
\icmlauthor{Brennan Abanades}{roche}
\icmlauthor{Robert G. Alberstein}{pd}
\icmlauthor{Pan Kessel}{pd}
\icmlauthor{Pranav Rao}{pd}
\icmlauthor{Jae Hyeon Lee}{pd}
\icmlauthor{Richard Bonneau}{pd}
\icmlauthor{Andrew M. Watkins}{pd}
\icmlauthor{Franziska Seeger}{pd}
\end{icmlauthorlist}

\icmlaffiliation{pd}{Prescient Design, Genentech, South San Francisco, CA, USA}
\icmlaffiliation{roche}{Large Molecule Research, Roche, Penzberg, Germany}

\icmlcorrespondingauthor{Fr\'ed\'eric A. Dreyer}{dreyer.frederic@gene.com}

\icmlkeywords{Machine Learning, ICML}

\vskip 0.3in
]

\printAffiliationsAndNotice{} 

\begin{abstract}
We introduce Ibex, a pan-immunoglobulin structure prediction model that achieves state-of-the-art accuracy in modeling the variable domains of antibodies, nanobodies, and T-cell receptors. 
Unlike previous approaches, Ibex explicitly distinguishes between bound and unbound protein conformations by training on labeled \textit{apo} and \textit{holo} structural pairs, enabling accurate prediction of both states at inference time. 
Using a comprehensive private dataset of high-resolution antibody structures, we demonstrate superior out-of-distribution performance compared to existing specialized and general protein structure prediction tools. 
Ibex combines the accuracy of cutting-edge models with significantly reduced computational requirements, providing a robust foundation for accelerating large molecule design and therapeutic development.
\end{abstract}

\section{Introduction}\label{sec:intro}
Protein structure plays a key role in understanding cellular function~\cite{kreitz2023programmable,lin2024comprehensive} and in structure-based drug or enzyme design~\cite{batool2019structure,isert2023structure,borkakoti2023alphafold2,borkakoti2025structural,dissanayake2025computational}.
Recent advances in machine learning have dramatically improved protein structure prediction from sequence, allowing for routine modeling of atomically accurate structures of novel complexes~\cite{af2,baek2021accurate,esmfold,af3}.
However, predicting alternative protein conformations, such as those adopted by fold-switching proteins~\cite{chakravarty2022alphafold2} or through protein-protein interactions, remains a significant challenge, particularly for conformations not well-represented in training data~\cite{schafer2025alphafold2}.

Modeling the full spectrum of protein conformational ensembles remains a grand challenge for computational biology. 
While physics-based approaches like molecular dynamics simulations can characterize these states~\cite{karplus2002molecular}, their accuracy is often constrained by approximate force fields and the prohibitive computational cost of sampling rare events, even with enhanced sampling techniques~\cite{bernardi2015enhanced}. 
To overcome these limitations, generative machine learning models have more recently been developed to emulate the equilibrium distribution of protein conformations, often by training on extensive simulated trajectories~\cite{jing2024alphafold,BioEmu2024}. 
For many applications in therapeutic design, however, resolving the full ensemble is not required. 
Instead, a more tractable and often sufficient goal is the accurate prediction of the distinct unbound and bound conformations. 
This is a task well-suited for machine learning models, which can be trained to predict these functionally critical states directly from sequence.

The accurate prediction of immune protein structures, most notably among them antibodies and T-cell receptors, is of critical importance to the design of better biologics and the acceleration of drug discovery~\cite{lu2020development,akbar2022progress}. 
Several dedicated specialized models have therefore emerged in recent years focusing on predicting the variable domain of antibodies~\cite{abanades2022ablooper,lee2022equifold,ruffolo2023fast,immunebuilder}.
These specialized approaches circumvent the computationally expensive multiple sequence alignment step typically required by general protein structure prediction models, as the complementarity determining region (CDR) of antibodies lacks the evolutionary conservation signals that multiple sequence alignments are designed to capture~\cite{di2007molecular,marks2012protein}.

\begin{figure*}
    \centering
    \hfill
    \includegraphics[width=0.36\linewidth]{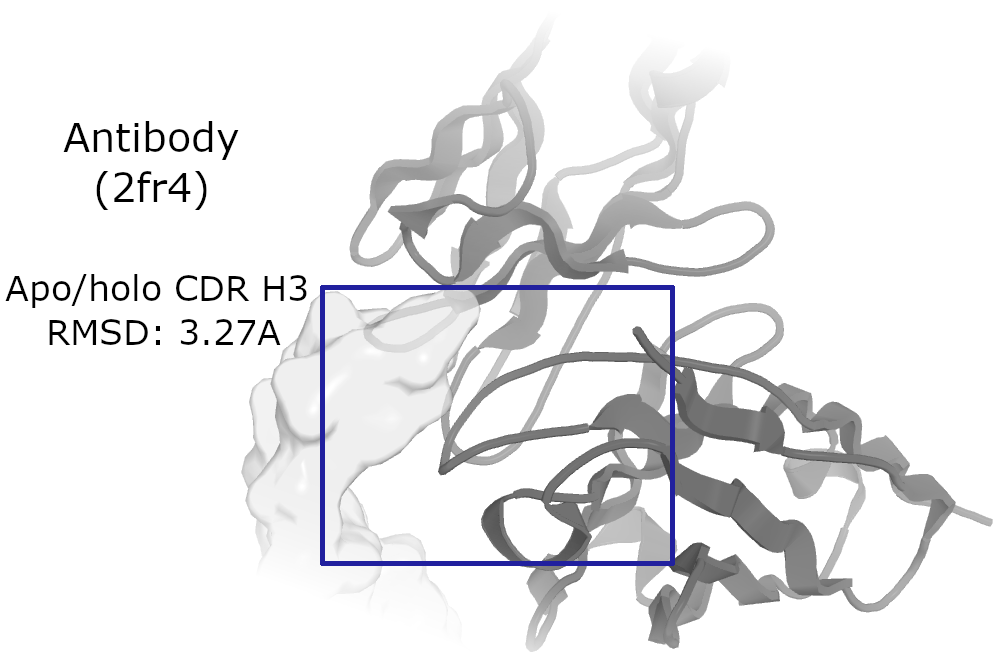}
    \hfill
    \includegraphics[width=0.35\linewidth]{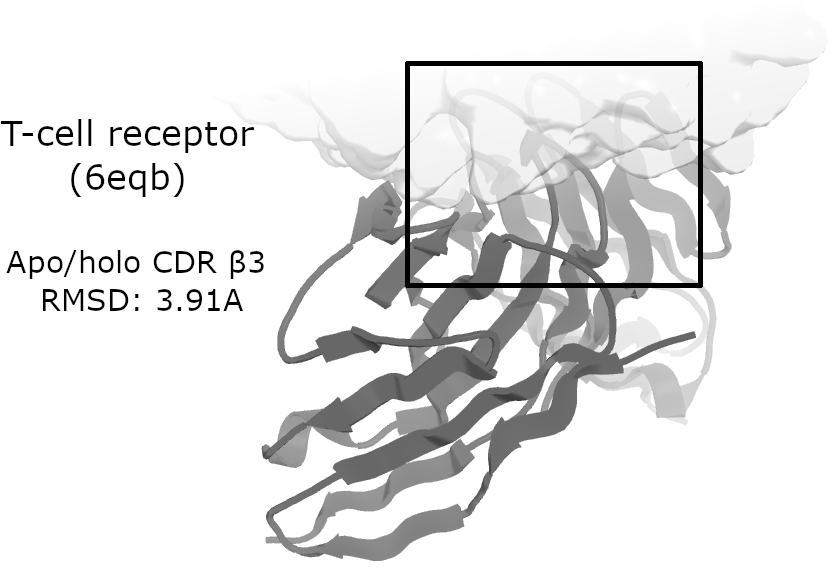}
    \hfill\vspace{2mm}\\
    \centering\includegraphics[width=0.247\linewidth,cfbox=darkblue 0.7pt 0.0pt]{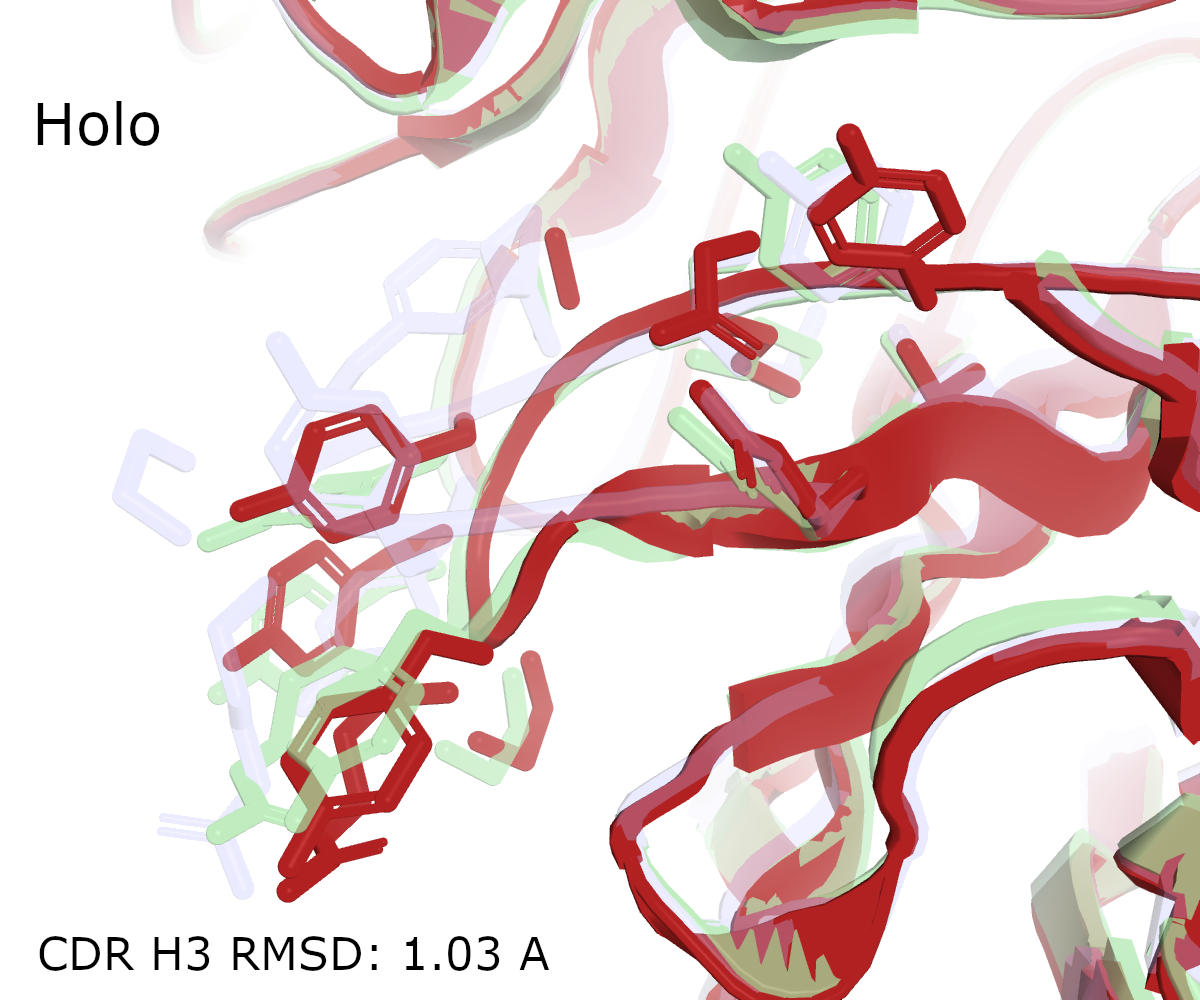}\hspace{-0.7pt}%
    \includegraphics[width=0.247\linewidth,cfbox=darkblue 0.7pt 0.0pt]{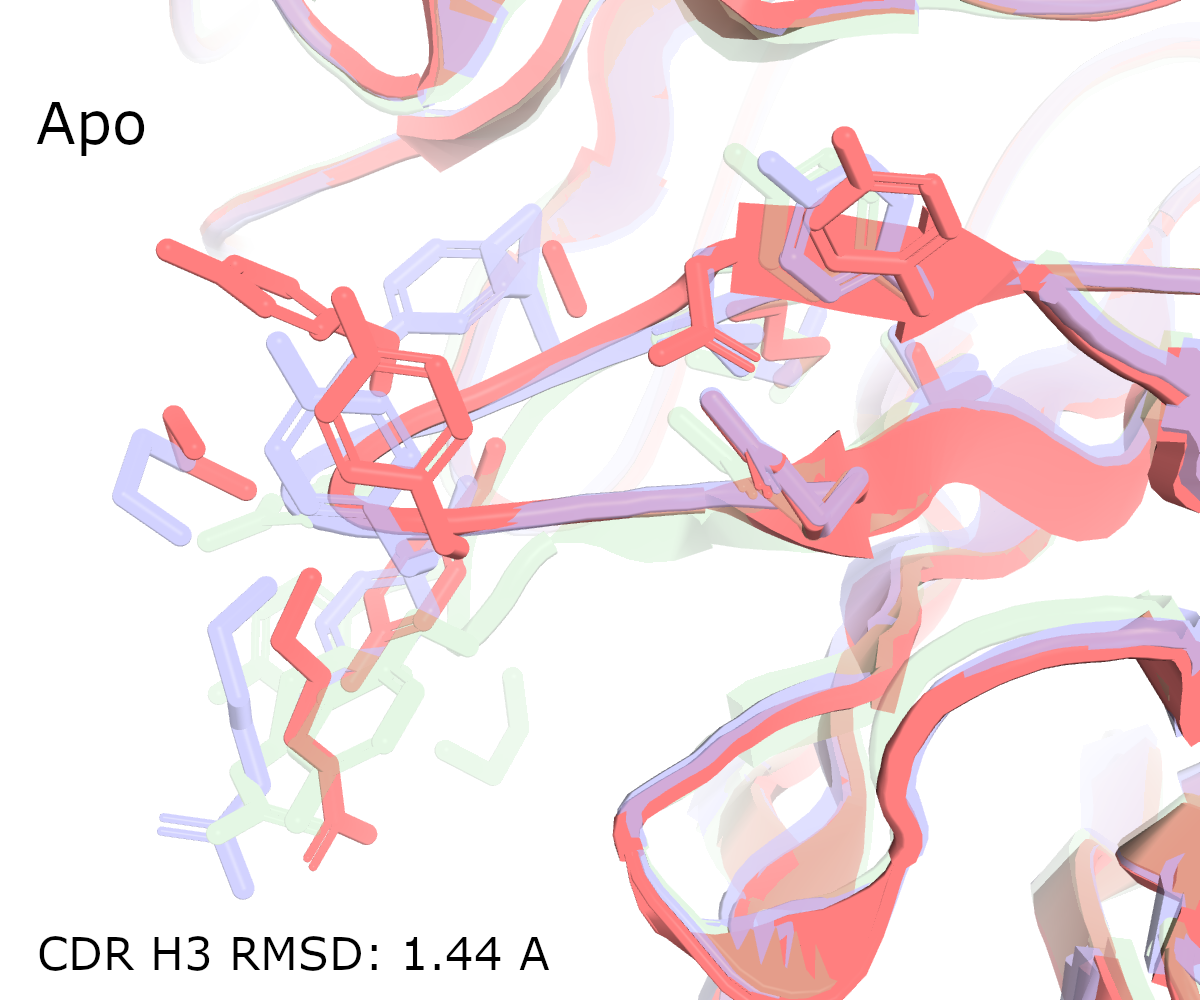}\hfill%
    \includegraphics[width=0.247\linewidth,cfbox=black 0.7pt 0.0pt]{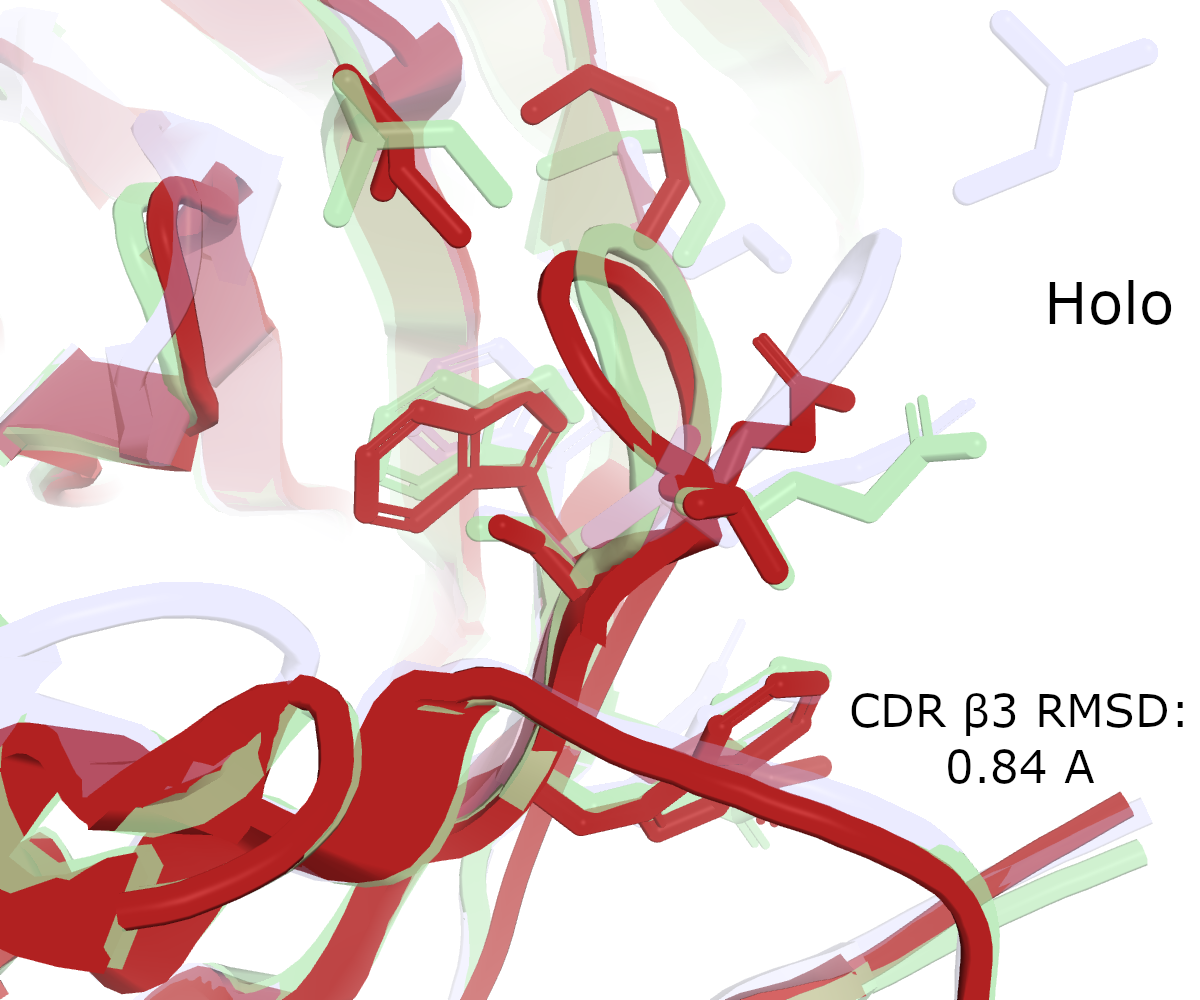}\hspace{-0.7pt}%
    \includegraphics[width=0.247\linewidth,cfbox=black 0.7pt 0.0pt]{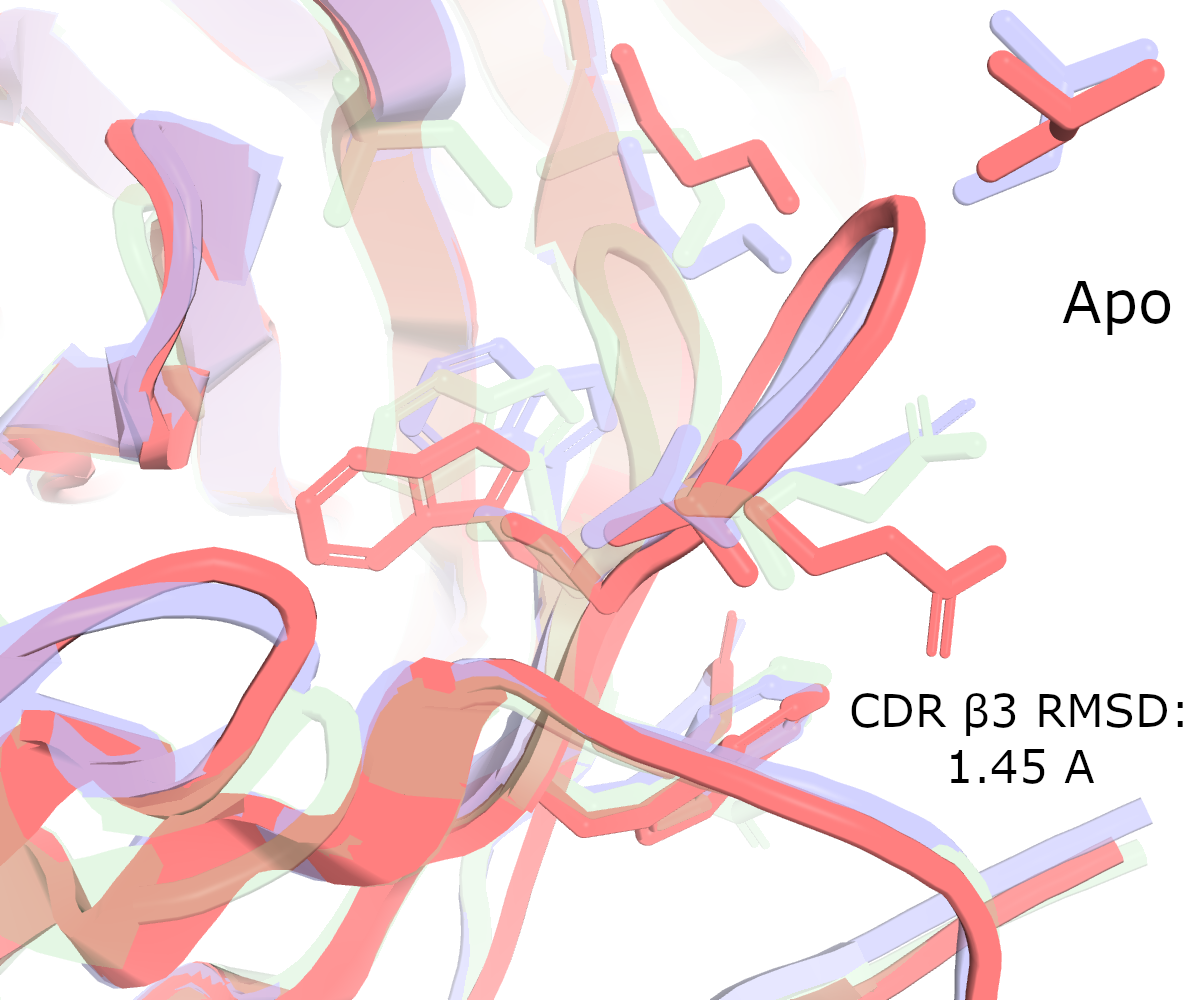}
    \caption{The first two panels show an antibody (PDB codes: \texttt{2fr4} and \texttt{1xf3}), with the \textit{holo} ground truth and predictions shown on the left, and \textit{apo} poses shown on the right.  The \textit{holo} Ibex predictions are in dark red and the \textit{apo} predictions in salmon, superimposed on their respective ground truth structures shown in green and blue. The view is centered on the H3 loop, with side chains shown for the loop residues. 
    The last two panels show the $\beta3$ loop of a T-cell receptor (\texttt{6eqb} and \texttt{4jfh}).}
    \label{fig:example}
\end{figure*}
    
A central task for antibody design is understanding the interactions between residues in the CDR loops and the antigen, particularly those involving the third CDR loop on the heavy chain (CDR H3)~\cite{NARCISO2011435}. 
The framework regions are relatively conserved, as are five of the six CDR loops, which tend to cluster under known canonical conformations~\cite{al1997standard}.
The CDR H3 loop, however, is uniquely diverse in sequence and length due to being defined by the recombination of three different genes, and can undergo substantial conformation change upon binding~\cite{GREENSHIELDSWATSON2025102983}.
Predicting this inherent conformational flexibility from sequence is a distinct challenge, with recent methods developed to classify CDR loops as either rigid or dynamic based on their local structural environment~\cite{spoendlin2025predicting}.
Improving modeling accuracy of the CDR H3 loop is central to advancing rational antibody design, as it often drives antigen recognition~\cite{tsuchiya2016diversity}, yet its structural diversity makes it particularly challenging to model accurately.

In this work, we address these challenges by introducing Ibex, a pan-immunoglobulin deep learning model for predicting the structure of antibodies, nanobodies, and TCRs. Ibex introduces two key advances: first, it is trained on a broad corpus of immunoglobulin and immunoglobulin-like domains to improve generalization, and second, it is explicitly trained to distinguish between binding states by using labeled \textit{apo} (unbound) and \textit{holo} (bound) structural pairs. 
This allows Ibex to generate predictions of both conformations from a single sequence, as illustrated in Figure~\ref{fig:example}.

We validate Ibex's performance on two distinct benchmarks: a standard test set of publicly available structures and a challenging internal test set of several hundred high-resolution, unpublished antibodies with novel CDR H3 loops. Our results show that Ibex achieves state-of-the-art accuracy across both benchmarks while requiring significantly less computational resources than recent competing models.

\begin{figure*}[t]
    \centering
    \includegraphics[width=0.34\linewidth]{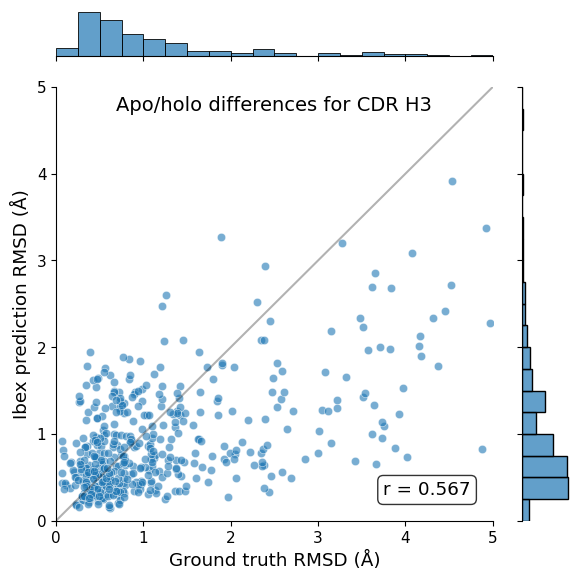}%
    \includegraphics[width=0.32\linewidth]{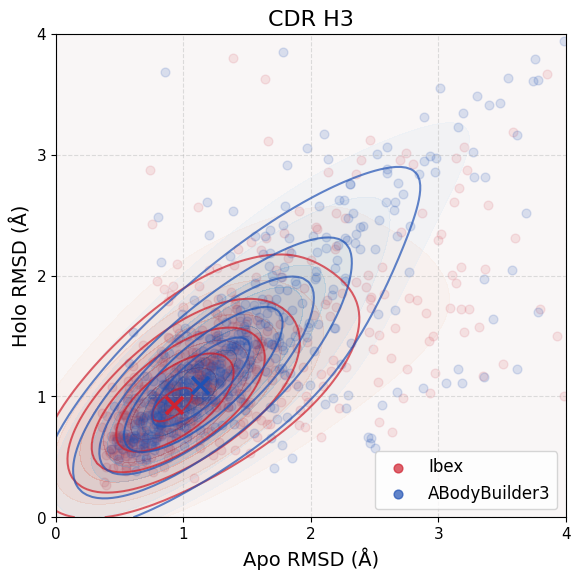}%
    \includegraphics[width=0.32\linewidth]{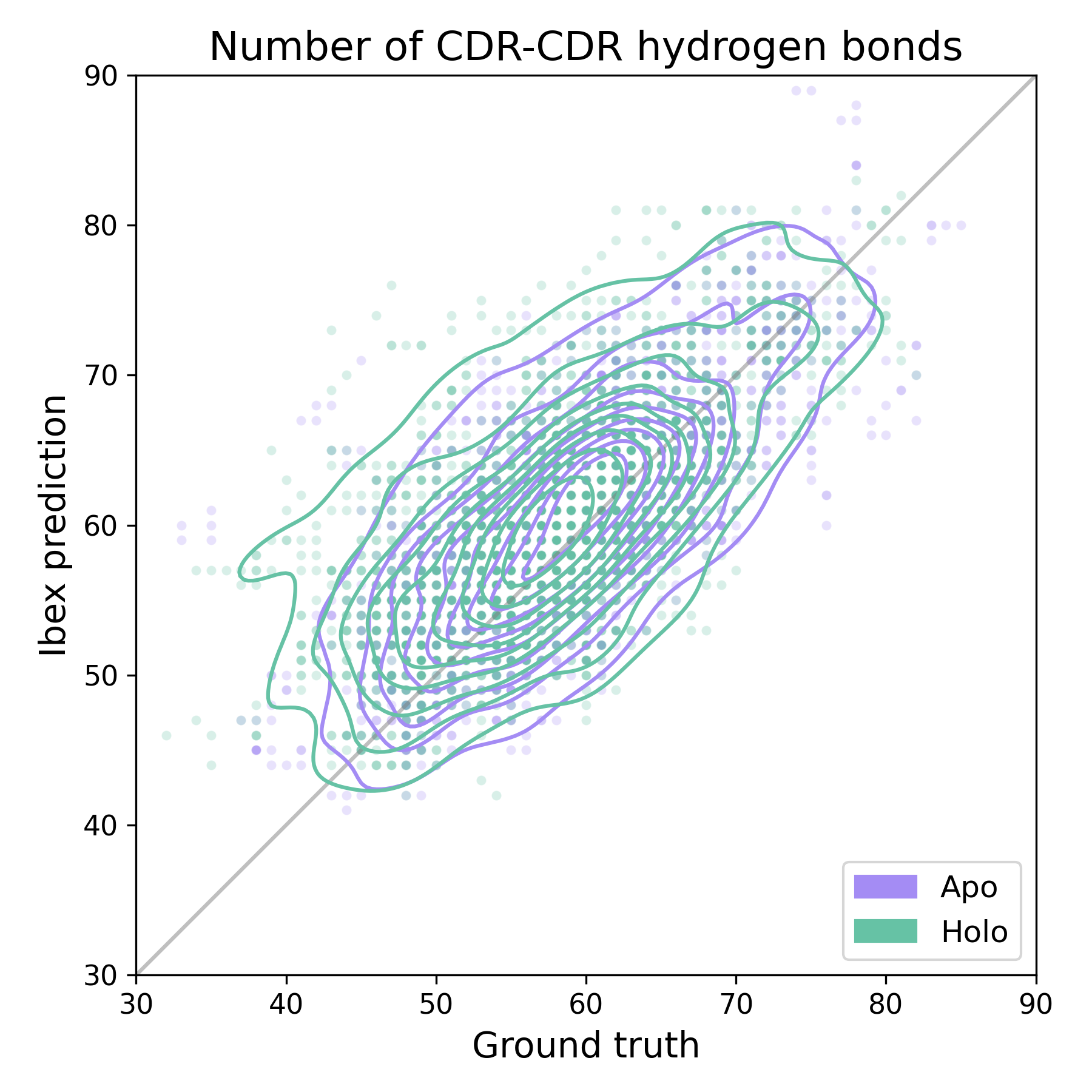}
    \caption{A. RMSD between matching \textit{apo} and \textit{holo} pairs, comparing ground truth structures to Ibex predictions.
    B. RMSD between predicted structures and their corresponding ground truth for matched \textit{apo} and \textit{holo} pairs. The mode of the distribution is indicated with a cross.
    C. Comparison of the number of hydrogen bonds in CDR-CDR residue interaction networks for matching \textit{apo} and \textit{holo} pairs.
    }
        \vspace*{-46ex}
        A
        \hspace{34ex}
        B
        \hspace{33ex}
        C
        \hspace{31ex}
        \vspace*{43ex}

    \label{fig:apo-holo}
\end{figure*}

\section{The Ibex Model}
\label{sec:ibex}
Ibex is a deep learning structure prediction model that builds upon the AlphaFold2 architecture~\cite{af2}. Its core consists of 16 structure module blocks that iteratively refine a 3D structure from sequence-based features. A key architectural innovation is the introduction of a conformation token, which is provided as an input feature alongside the variable region sequence. 
During training, this token explicitly labels structures as either \textit{apo} or \textit{holo}, enabling the model to learn the distinct structural features of each state. At inference, this token can be specified to directly generate a prediction for the desired conformational state from an antibody sequence. To ensure this state information effectively propagates through the network, a residual connection was added from the initial embedding to every structure module block.

The model was trained using a curriculum learning strategy~\cite{curriculum} designed to maximize accuracy and generalization. The training data combines three distinct sources: (1) a high-quality dataset of 14k experimental antibody, nanobody, and TCR structures from SAbDab~\cite{sabdab} and STCRDab~\cite{stcrdab}, annotated as apo or holo; (2) a broader dataset of 22k immunoglobulin-like domains from the Protein Data Bank~\cite{pdb}; and (3) a large distillation set of over 60k diverse, predicted antibody structures. The three-stage training curriculum starts with pre-training on the combined dataset, then specializes on high-quality experimental data before fine-tuning using a structural violation loss. The final Ibex model is an ensemble of eight independently trained models, which returns the prediction closest to the mean of the ensemble. Further details on the model architecture and training strategy are given in Section~\ref{sec:methods}.

\section{Results}
\label{sec:results}

We evaluate the performance of Ibex by first assessing its capability to model the conformational transition between apo and holo states. We then benchmark its accuracy against state-of-the-art models on a public test set and finally test its generalization performance on a large, private dataset of high-resolution antibody structures.

\subsection{Analysis of paired \textit{apo}/\textit{holo} structures}
\label{sec:apo-holo}

A key innovation of Ibex is its ability to recapitulate distinct conformational states of antibodies. From a single sequence, one can thus predict both the bound and unbound forms. 
To validate this, we analyze the conformational changes between 562 experimentally determined \textit{apo}/\textit{holo} antibody pairs and compare them to the changes predicted by Ibex.

In Figure~\ref{fig:apo-holo}A, we show the root mean square deviation (RMSD) in angstroms of the CDR H3 loop between apo and holo structures, comparing Ibex predictions to the ground truth. Here the loop RMSD is computed by averaging over backbone atoms after aligning the heavy chain according to framework residues. Each point corresponds to a matching pair of experimentally determined structures with both conformations for the same variable domain sequence.
We observe that Ibex has a reasonable correlation with experimentally observed differences, and produces a similar distribution of overall conformational change. However, the model sometimes underestimates conformation changes, particularly in the case of outliers with very large observed RMSD values. It is important to note here that most of the known paired \textit{apo}/\textit{holo} structures were included in the training.

\begin{table*}
    \centering
    \begin{tabular}{c@{\hspace{1em}}c|cccccccc}
        & & CDR H1 & CDR H2 & CDR H3 & Fw H & CDR L1 & CDR L2 & CDR L3 & Fw L \\
        \midrule
        \multirow{5}{*}{\rotatebox[origin=c]{90}{Antibodies}} 
        & ESMFold & 0.70 & 0.99 & 3.15 & 0.65 & 1.04 & 0.50 & 1.29 & 0.61 \\
        & ABodyBuilder3 & 0.71 & 0.65 & 2.86 & 0.51 & 0.67 & 0.49 & 1.13 & 0.59\\
        & Chai-1 & 0.67 & \textit{0.53} & \textbf{2.65} & \textbf{0.45} & \textbf{0.53} & \textit{0.41} & 1.20 & 0.54 \\
        & Boltz-1 & \textit{0.63} & \textbf{0.52} & 2.96 & 0.47 & \textit{0.54} & \textbf{0.38} & \textbf{0.96} & \textbf{0.52} \\
        & Ibex & \textbf{0.61} & 0.57 & \textit{2.72} & \textbf{0.45} & 0.57 & 0.43 & \textit{0.98} & \textbf{0.52} \\
        \midrule
        \multirow{5}{*}{\rotatebox[origin=c]{90}{Nanobodies}} 
        & ESMFold & \textit{1.55} & \textbf{0.94} & 3.60 & 0.63 & - & - & - & - \\
        & NanoBuilder2 & 1.74 & 1.22 & 3.31 & 0.76 & - & - & - & - \\
        & Chai-1 & \textit{1.57} & 1.17 & 3.76 & 0.64 & - & - & - & - \\
        & Boltz-1 & 1.59 & 0.97 & \textbf{2.83} & \textbf{0.61} & - & - & - & - \\
        & Ibex & 1.62 & \textit{0.96} & \textit{3.12} & \textit{0.62} & - & - & - & - \\
        \midrule
        & & CDR $\beta$1 & CDR $\beta$2 & CDR $\beta$3 & Fw $\beta$ & CDR $\alpha$1 & CDR $\alpha$2 & CDR $\alpha$3 & Fw $\alpha$ \\
        \midrule
        \multirow{5}{*}{\rotatebox[origin=c]{90}{TCRs}} 
        & ESMFold & 0.68 & 0.66 & 2.49 & 0.71 & 1.35 & 0.96 & 2.31 &  0.75 \\
        & TCRBuilder2+ & 0.72 & 0.79 & \textit{1.85} & 0.68 & 1.18 & 1.04 & \textit{2.00} & 0.94 \\
        & Chai-1 & 0.60 & \textbf{0.52} & 2.10 & \textit{0.58} & 1.03 & 0.89 & 2.13 & \textit{0.66} \\
        & Boltz-1 & \textit{0.59} & \textit{0.53} & 2.40 & 0.66 & \textbf{0.95} & \textit{0.87} & 2.05 & \textbf{0.58} \\
        & Ibex & \textbf{0.57} & 0.57 & \textbf{1.84} & \textbf{0.60} & \textit{1.02} & \textbf{0.85} & \textbf{1.93} & 0.71 \\
    \end{tabular}
    \caption{Average RMSD in angstrom, evaluated separately for each region on the ImmuneBuilder test set of antibodies, nanobodies and TCRs.}
    \label{tab:results}
\end{table*}

Having evaluated comparative differences between experimental and predicted conformational changes, we now evaluate the performance of Ibex at recapitulating the respective ground truth structure. To this end, we show in Figure~\ref{fig:apo-holo}B the Ibex CDR H3 RMSD across all matching \textit{apo}/\textit{holo} antibody pairs. The $x$ coordinate gives the RMSD between predicted and experimental \textit{apo} structures, while the $y$ coordinate gives the corresponding \textit{holo} RMSD. 
Here we compare Ibex against ABodyBuilder3~\cite{abb3}, a recent antibody structure prediction model. While each datapoint for Ibex corresponds to two distinct predictions, ABodyBuilder3 can only provide a single conformation for a specified sequence, and we use the same prediction to evaluate the RMSD to the \textit{apo} and \textit{holo} ground truth in that case. 
We show contours of the kernel density estimation of the joint probability distribution function, with the mode of the distribution indicated by a cross.
Ibex achieves markedly better performance, with a narrower distribution that has fewer large RMSD outliers.

Finally, we consider whether Ibex's predictions reflect the corresponding biophysical states associated with conformational changes by quantifying the recapitulation of hydrogen-bond networks within the CDRs using the Rosetta ScoreFunction~\cite{alford2017rosetta}. 
Figure~\ref{fig:apo-holo}C shows that Ibex adequately recapitulates the expected number of hydrogen bonds in both \textit{apo} and \textit{holo} conformations, with a slight bias towards overproducing hydrogen bonds in the \textit{holo} state. between the number of hydrogen bonds observed in the ground truth structure compared to corresponding Ibex predictions.
Here predictions are evaluated on the same set of 562 matching \textit{apo}/\textit{holo} pairs.
As detailed in Appendix~\ref{app:hydrogen}, precision and recall analysis of the predicted inter-CDR hydrogen bond contacts confirms the fidelity of these reproduced networks, indicating that Ibex correctly captures both the magnitude and connectivity of the intramolecular interactions stabilizing each conformation.

\subsection{Benchmarking on the ImmuneBuilder test set}
\label{sec:public-benchmark}

To assess its performance against existing methods, we benchmark Ibex on a standardized test set derived from the ImmuneBuilder suite~\cite{immunebuilder}, which includes antibodies, nanobodies, and TCRs. 
For consistent benchmarking, we combined the respective test sets and ensured that no structures from a cluster containing a test set member were used during training. 
We compare against ESMFold~\cite{esmfold}, Chai-1~\cite{chai} and Boltz-1~\cite{boltz}, which are general protein structure prediction models, as well as against ABodyBuilder3, NanoBuilder2 and TCRBuilder2+~\cite{tcrbuilder2p}, three specialized models for antibodies, nanobodies and TCRs respectively.
For Chai-1 and Boltz-1, a single seed and diffusion trajectory is used in all comparisons.
We do not include Boltz-2~\cite{boltz2} in this benchmark, as it uses a later PDB cutoff date for the curation of its training data, and thus was trained on most of the ImmuneBuilder test set.
Furthermore, to mitigate data leakage, we removed six structures from the nanobody test set for which a matching CDR H3 loop sequence was identified in the NanoBody2 training or validation splits.
The RMSD is evaluated separately for each region by aligning each chain independently based on their framework residues and averaging over corresponding backbone atom RMSD.

As summarized in Table \ref{tab:results}, Ibex demonstrates state-of-the-art or highly competitive performance across all immune protein types and regions. Notably, Ibex achieves the lowest mean RMSD for the CDR $\beta$3 and CDR $\alpha$3 TCR loops, and the second lowest for both antibody and nanobody CDR H3 loops.
Chai-1 achieves marginally better results on the CDR H3 of antibodies, but poorly models the antibody CDR L3 loop as well as nanobodies and TCRs. 
Boltz-1, meanwhile, achieves the best results on the CDR H3 loop of nanobodies, but performs worse on the antibody CDR H3 loop and the TCR CDR $\beta$3 loop. 
Ibex largely surpasses the performance of specialized models across all modalities. More detailed comparisons provided in Appendix~\ref{app:comparisons}.

\begin{figure*}
    \centering
    \begin{minipage}{0.68\linewidth}
        \centering
        \includegraphics[width=\linewidth]{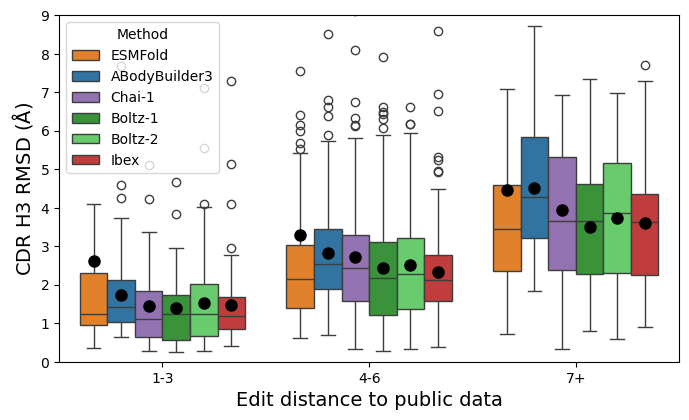}
    \end{minipage}%
    \hfill%
    \begin{minipage}{0.28\linewidth}
        \centering
        \vspace{-1.5mm}
        \begin{tabular}{l c}
            \toprule
            \multicolumn{2}{l}{\textbf{Mean CDR H3 RMSD (\AA{})}}\\            \midrule
            ESMFold        & 3.27 \\
            ABodyBuilder3  & 2.78 \\
            Chai-1         & 2.55 \\
            Boltz-1        & \textit{2.30} \\
            Boltz-2        & 2.42 \\
            Ibex           & \textbf{2.28} \\
            \bottomrule
        \end{tabular}
    \end{minipage}
    \caption{Benchmark on a private dataset of 286 antibody structures, showing the CDR H3 RMSD as a function of edit distance to the closest matching H3 loop in SAbDAb. The box shows the lower and upper quartile, with the median represented as a horizontal line and the mean as a filled circle. Outliers are shown as empty circles, with ESMFold having outliers outside of the plotted range.}
    \label{fig:internal_benchmark}
\end{figure*}

\subsection{Benchmarking on a private dataset}
\label{sec:private-benchmark}

Finally, we evaluate Ibex on a curated private dataset of 286 unique and novel antibody structures. Each structure in this dataset has a unique CDR H3 loop which differs by at least one amino acid from any known public antibody structure, and 45 have an edit distance of 7 or more from any CDR H3 loop in SAbDab. The length of the CDR H3 loops of these structures ranges from 5 to 22, with an average experimental resolution of 2.1\AA{}. Further specifics on the dataset composition are given in Appendix~\ref{app:private-dataset}.

We show the CDR H3 RMSD on this dataset in Figure~\ref{fig:internal_benchmark} as a function of edit distance to the closest matching CDR H3 loop in SAbDab. 
We observe that while ABodyBuilder3 achieves high performance on the test set, it is not as robust out-of-distribution compared to state-of-the-art general protein structure prediction models.
In contrast, Ibex shows comparable performance to Boltz-1, which may be attributed to distillation from predicted structures and the use of a broader corpus of training data.
We note that it has recently been shown that including antigen context can sometimes improve CDR H3 modeling accuracy in co-folding models~\cite{hitawala2025does}, which is not provided here.
It is also interesting to note that ESMFold achieves comparable median performance to more recent models in the high edit distance region, indicating that there has been limited progress in out-of-distribution robustness of CDR loop modeling. However, ESMFold sometimes misfolds the variable region entirely, leading to high RMSD outliers and a larger average RMSD value compared to other methods.
A comparison of inference compute time is provided in Appendix~\ref{app:compute}.

\begin{figure}
    \centering
    \includegraphics[width=1.0\linewidth]{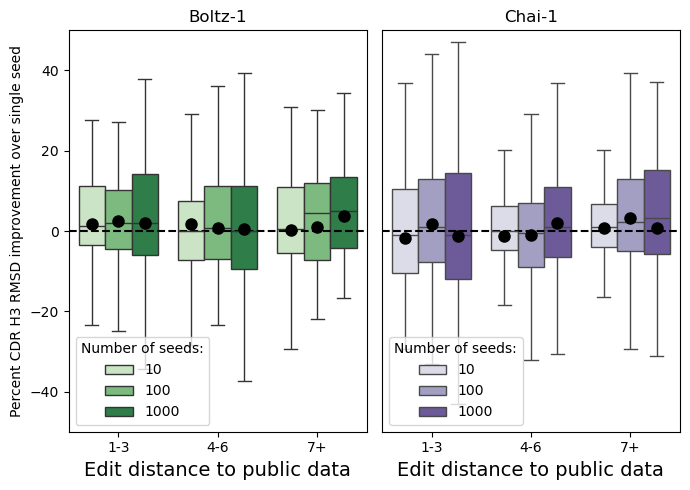}
    \caption{Relative improvement in CDR H3 RMSD from considering the best scoring prediction out of 10, 100 and 1000 seeds for Boltz-1 and Chai-1, evaluated on a private dataset of 286 antibodies and shown as a function of edit distance to the closest matching H3 loop in SAbDab.
    In each case, the best scoring prediction from the corresponding range of seeds is compared to a single seed baseline.}
    \label{fig:scaling}
\end{figure}

Due to their stochastic nature, diffusion-based structure prediction models such as AlphaFold3 can be sampled repeatedly, ranking results with their own confidence head.
Notably, recent studies have shown that sufficient sampling can dramatically improve quality of antibody-antigen docking poses generated by the model~\cite{af3,xu2025foldbench}.
We therefore aim to investigate whether similar scaling behavior could be observed in the modeling accuracy of the CDR H3 loop.
In Figure~\ref{fig:scaling}, we evaluate Boltz-1 and Chai-1 on our private dataset, sampling up to 1000 structures for each sequence, using distinct seeds and a single diffusion trajectory per seed.
For each sequence, we then evaluate the CDR H3 RMSD between the highest confidence prediction and a single seed prediction.
In contrast to previous docking experiments, we observe very little improvement in prediction accuracy when scaling to 1000 seeds. 
Both models also mostly do not predict varying conformations of the loop backbone for the same sequence, with RMSD between predictions generally varying by only a few percent.
This might be due to scarcity of \textit{apo} data in SAbDab, and because very few antibodies are resolved in more than one conformation.
It is also possible that the confidence score is not as effective at evaluating unbound protein loops as it is at scoring protein-protein interfaces.

\section{Discussion}
\label{ref:discussion}

The prediction of antibody conformational changes upon antigen binding is a central challenge in computational biology, directly impacting our ability to design and engineer effective therapeutics. In this work, we have demonstrated that by explicitly conditioning a deep learning model, Ibex, on the binding state, we can generate distinct and accurate structures for both the \textit{apo} and \textit{holo} conformations from a single variable region sequence. This capability, combined with state-of-the-art performance on out-of-distribution antibody sequences, represents a significant advance toward the \textit{in silico} design and optimization of next-generation biologics.

A key innovation of Ibex is its ability to resolve part of the inherent ambiguity that arises when training on structural databases containing multiple entries for the same sequence. Previous approaches, trained on undifferentiated data, risk predicting a non-physical average or collapsing to the most common conformational state due to an ambiguous minimization objective. 
Through the introduction of a conformation token, Ibex learns to predict distinct state-specific minima in the energy landscape and can recapitulate known \textit{apo}/\textit{holo} conformational changes. 
This has practical implications, enabling improved property prediction for antibodies that display induced-fit recognition~\cite{rini1992structural} and more accurate docking through the selection of the correct starting point conformation.

Leveraging a private dataset of hundreds of antibodies with novel CDR H3 loops, we demonstrate that Ibex achieves superior out-of-distribution generalization compared to both specialized and general protein structure prediction models. This provides a critical test for models intended for real-world therapeutic design, where sequences are inherently novel. We attribute this performance to our comprehensive training strategy, which integrates a broad corpus of immunoglobulin-like domains and distillation from predicted structures to a canonical dataset of experimental antibody and TCR structures.

Despite these advances, several limitations remain. The binary classification of protein states into apo and holo is a useful but significant simplification of a complex conformational landscape~\cite{marks2018predicting}. 
Ibex will thus be unable to recapitulate experimental data with multiple distinct bound, or unbound, antibody crystal structures~\cite{garces2020molecular}.
Furthermore, the paucity of experimental data containing large conformational changes constrains Ibex's capabilities in capturing large structural rearrangements and in generalizing to novel conformational shifts.

Our work opens however several exciting avenues for further research. An immediate next step is to incorporate antigen structural information, allowing the model to predict binding-induced conformational changes in a target-specific manner. This would pave the way for high-fidelity \textit{in silico} affinity prediction. Moreover, the conformation token is a generalizable concept that could be expanded to model other biological states, such as those dependent on pH or allosteric modulation. Finally, part of the experimental data scarcity could be bridged by augmenting training sets with extensive conformational ensembles generated from molecular dynamics simulations, and by broadening the immunoglobulin-like dataset to incorporate relevant loop-mediated protein-protein interfaces~\cite{kovtun2024pinder}.

In conclusion, Ibex provides a fast, accurate, and conformational-aware tool for immunoglobulin structure prediction. By bridging the gap between static and dynamic views of antibody binding and demonstrating robust generalization to novel sequences, it lays a critical foundation for accelerating the computational design, screening, and engineering of antibody-based therapeutics.

\section{Methods}
\label{sec:methods}

\subsection{Model}
\begin{figure}
    \centering
    \includegraphics[width=1.0\linewidth]{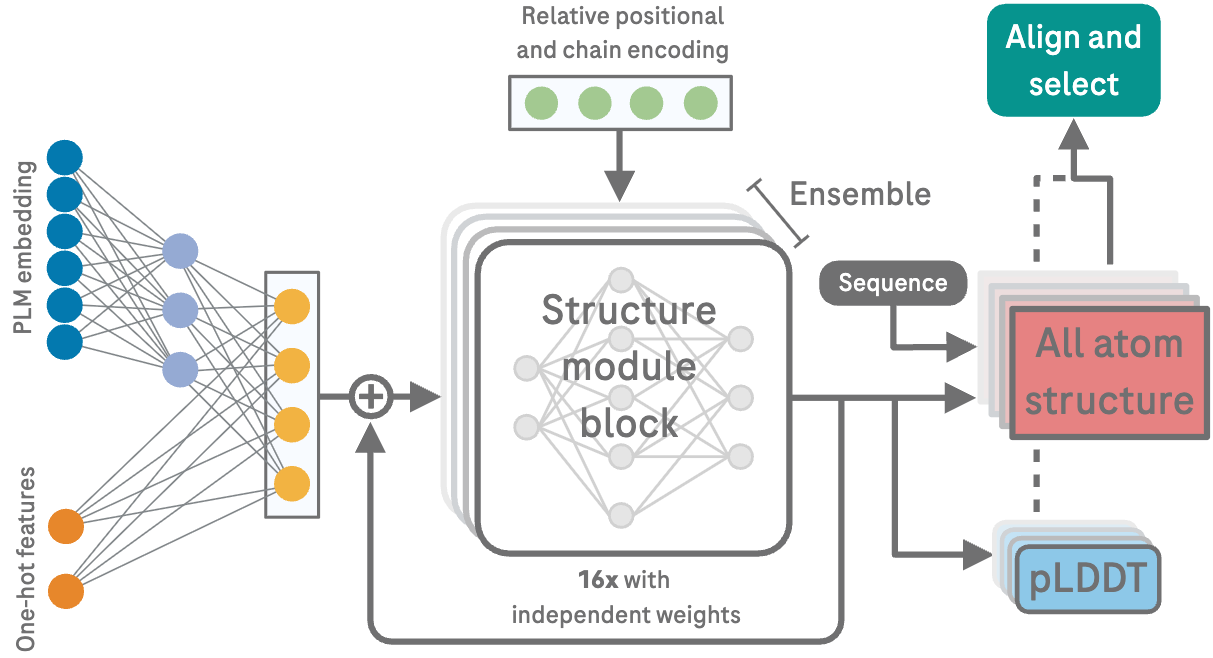}
    \caption{Overview of the Ibex model. The one-hot encoded residue features are concatenated with a projection of language model embeddings and used as input to 16 structure module blocks. The final residue representation is then used to predict all atomic coordinates and uncertainties. }
    \label{fig:architecture}
\end{figure}

Ibex follows a similar architecture to ABodyBuilder2~\cite{immunebuilder} and uses several structure module blocks from AlphaFold2~\cite{af2} to refine structures from an input sequence embedding. The model implementation leverages code from OpenFold~\cite{openfold} and ABodyBuilder3~\cite{abb3}.
An overview of the model is given in Figure~\ref{fig:architecture}.

We use as input a combination of ESM-C 300M sequence embeddings~\cite{esmc}, the one-hot encoding of the sequence, and a residue-level encoding of the chain and conformation feature. 
The conformation token is provided as a label during training, and used as an additional input at inference, to disambiguate between \textit{apo} and \textit{holo} structures.
The language model embeddings are passed to a two-layer MLP and concatenated with the remaining node features after applying a layer normalization.
Pairwise edge features are set to the one-hot encoding of relative positions in the sequence, concatenated with an encoding of which chains both residues belong to. For each invariant point attention layer in the structure module, this pair representation is concatenated with a distance feature map and provided as input.
The core of the model consists of 16 consecutive and independent structure module blocks, which update residue coordinates and single representation through an invariant point attention layer. 
A key change from ABodyBuilder3 is the introduction of a residual connection between the initial residue representation embedding and each structure module block, which improves training convergence and ensures information from the conformation token is preserved throughout the network.
The residue representation from the last structure module block is used to predict backbone atom coordinates.
The original input sequence is then used to reconstruct side-chain atoms from idealized coordinates using predicted chi-angles.
Uncertainties are modeled through a predicted local-distance difference test (pLDDT) head, which predicts a projection of local confidence into 50 bins.

\subsection{Data}
Ibex is intended to be a pan-immunoglobulin structure prediction model, and we therefore aim to collate a comprehensive dataset of all available immunoglobulin protein structures, with explicit labeling of their binding state.
The training data for Ibex is constructed from three different sources, which are clustered and combined dynamically throughout the training process.
Each structure is also annotated with a conformation token, which labels them as \textit{apo} or \textit{holo} based on the presence of an antigen to which the variable region is bound.
The model implicitly learns the distinctive structural motifs and folding patterns that differentiate antibodies, TCRs, and nanobodies.

The first source is a dataset of structures of antibody variable regions, nanobodies and TCR variable regions. These are curated from the structural antibody database (SAbDab)~\cite{sabdab} and the structural TCR database (STCRDab)~\cite{stcrdab}. 
Sequences are numbered using Anarci~\cite{anarci}, and we use the North definition~\cite{north} to delineate the CDR residues.
Structures are characterized as \textit{apo} if no antigen chain is indicated in the SAbDab or STCRDab metadata, otherwise they are labeled as \textit{holo}.
We remove structures with a resolution above 3.5\AA{}, as well as structures with a CDRH3 or $\beta$3 loop of more than 35 residues. We also remove all antibody variable domains for which one of the six Abangle~\cite{abangle} VH-VL orientation angle or distance values are more than five standard deviations from the mean computed over SAbDab. 
We are left with 14k structures, 760 of which are matched apo/\textit{holo} pairs for which both bound and unbound conformations are known. All structures are clustered based on the sequence of their concatenated CDR loops, using MMseqs2~\cite{mmseqs2,mmseqs2gpu} with 95\% sequence identity, resulting in 4.2k unique clusters.

Our second dataset consists of immunoglobulin-like domains found in the PDB. We identify any immunoglobulin-like structures according to the ECOD~\cite{cheng2014ecod} X-groups "Immunoglobulin-like beta-sandwich" and "jelly-roll" with PDB codes absent from both SAbDab and STCRDab. 
These are processed as individual chains, and loop residues are defined with the DSSP algorithm~\cite{dssp}. We assign apo labels to chains for which no heavy atoms are found within 10\AA{} of a loop residue.
We obtain a dataset of 22k single domain structures with resolution below 3.5\AA{}, which are clustered at 80\% sequence identity into 3k unique clusters.

Finally, we curate a dataset of predicted structures from paired sequences from the Observed Antibody Space (OAS)~\cite{oas1,oas2} using ESMFold~\cite{esmfold} and Boltz-1~\cite{boltz}.
Starting from 1.7 million unique paired sequences from OAS, we cluster them using concatenated CDR loops with a 60\% sequence identity threshold, as well as on the H3 sequence with 50\% minimum sequence identity. We then consider all cluster representatives from the concatenated CDR clustering which are also in distinct H3 clusters, resulting in a highly diversified set of 91k paired sequences.
Using ESMFold, we predict the corresponding structures, and filter out any for which one of the Abangle value is more than three standard deviations from the SAbDab mean, leaving us with 51k ESMFold structures. We randomly sample 10k of the 40k sequences for which ESMfold structures failed our Abangle filter, and predict their structure with Boltz-1, which are again filtered through the same procedure, resulting in 9k accepted Boltz-1 structures. We perform a short relaxation with OpenMM of all predicted structures.
All predicted structures are labeled as unbound.
This provides us with a dataset of 60k data points with very high CDR diversity which we use for distillation and to improve generalization. 

Ablation studies on the impact of training dataset size and of the different data sources are provided in Appendix~\ref{app:ablations}.

\subsection{Training}
We train Ibex in three stages, using a curriculum learning strategy~\cite{curriculum}, where training is slowly specialized towards experimental antibody and TCR data after pre-training on a larger corpus of predicted structures and related experimental protein data.

The first two stages use a Frame Aligned Point Error (FAPE) loss along with a backbone torsion angle and plDDT losses. The FAPE loss is clamped at 10\AA{}, and 30\AA{} when it is computed between CDR and framework residues, or between loop and non-loop annotated residues in the case of immunoglobulin-like single domains. 
Following a similar approach to AlphaFold2~\cite{af2}, the total loss term is the sum of the average backbone FAPE loss across each layer, the full atom FAPE loss from the final structure, as well as side-chain and backbone torsion angle losses and a pLDDT loss. The pLDDT loss consists of a cross-entropy loss on the discretised per-residue lDDT-$C\alpha$.

The third stage introduces structural violation losses. These follow AlphaFold2-Multimer~\cite{af2m} and penalize bond length violations, bond angle violations and steric clashes of non-bonded atoms.

Across each stage, we use an AdamW optimizer~\cite{adamW}. The first stage is trained with a constant learning rate of $2\cdot 10^{-4}$ for 3000 epochs. 
The second stage lasts 2000 epochs, and starts with a learning rate of $2\cdot 10^{-4}$ and slowly anneals to $2\cdot 10^{-5}$ from 200 epochs onwards using a lambda scheduler.
The third stage is trained with a constant learning rate of $5\cdot 10^{-5}$ for 1000 epochs.

An epoch consists of randomly sampling $N$ structures according to weighted probabilities assigned to each data point, where $N$ is set to the total size of the SAbDab and STCRDab structures used. 
We assign a probability inversely proportional to the cluster size to each data point to improve out-of-distribution generalization \cite{loukas2024generalizing}.
Matched pairs of \textit{apo}/\textit{holo} structures are identified and sampled separately from the unpaired SAbDAb and STCRDab structures to ensure they appear together in a batch. We upsample nanobodies and TCR by a factor 1.5, while paired \textit{apo}/\textit{holo} are upsampled by a factor 2.
In the first stage, the weighting of different data sources is split as 30\% SAbdab and STCRDab, 40\% predicted structures, and 30\% from the broader immunoglobulin protein dataset. This is changed to 70\% SAbDAb and STCRDab, 20\% predicted structures and 10\% immunoglobulin proteins in the second stage. 
Finally, in the third stage, we use 95\% of SAbDab and STCRDab structures and 5\% of predicted structures, this time restricting only to Boltz-1 samples and discarding the ESMFold and broader immunoglobulin data.

The final Ibex model consists of an ensemble of eight independently trained models. Predictions of all eight models are aligned, and the model returns the prediction closest to the mean.
Each model in the ensemble is trained using a different validation set of 150 antibody structures randomly selected from the SAbDab clusters. 
The total training time for all eight models in the ensemble was about 10 days on 64 H100 NVIDIA GPUs.

\section*{Data availability}
Data used to train Ibex was extracted from the public databases SAbDab~\cite{sabdab}, STCRDab~\cite{stcrdab} and PDB~\cite{pdb}.
Inference code is made available at \url{https://github.com/prescient-design/ibex}.
Model weights can be obtained at \url{https://zenodo.org/records/15866555}~\cite{zenodo}.

\section*{Acknowledgements}
We are grateful to David Konerding, Joseph Kleinhenz, and Henri Dwyer for engineering support, and to the Structural Biology teams at Genentech and Roche for providing structures used in benchmarking. 
We thank Henry Kenlay, Charlotte Deane, Allen Goodman, Josh Southern, Andrew Leaver-Fay, Nathan Frey, Lian Huang, as well as the entire Prescient Design team for useful discussions.

\bibliography{references}

\begin{thebibliography}{61}
\providecommand{\natexlab}[1]{#1}
\providecommand{\url}[1]{\texttt{#1}}
\expandafter\ifx\csname urlstyle\endcsname\relax
  \providecommand{\doi}[1]{doi: #1}\else
  \providecommand{\doi}{doi: \begingroup \urlstyle{rm}\Url}\fi

\bibitem[Abanades et~al.(2022)Abanades, Georges, Bujotzek, and Deane]{abanades2022ablooper}
Abanades, B., Georges, G., Bujotzek, A., and Deane, C.~M.
\newblock Ablooper: fast accurate antibody cdr loop structure prediction with accuracy estimation.
\newblock \emph{Bioinformatics}, 38\penalty0 (7):\penalty0 1877--1880, 2022.

\bibitem[Abanades et~al.(2023)Abanades, Wong, Boyles, Georges, Bujotzek, and Deane]{immunebuilder}
Abanades, B., Wong, W.~K., Boyles, F., Georges, G., Bujotzek, A., and Deane, C.~M.
\newblock Immunebuilder: Deep-learning models for predicting the structures of immune proteins.
\newblock \emph{Communications Biology}, 6\penalty0 (1):\penalty0 575, 2023.

\bibitem[Abramson et~al.(2024)Abramson, Adler, Dunger, Evans, Green, Pritzel, Ronneberger, Willmore, Ballard, Bambrick, Bodenstein, Evans, Hung, O'Neill, Reiman, Tunyasuvunakool, Wu, {\v Z}emgulyt{\.e}, Arvaniti, Beattie, Bertolli, Bridgland, Cherepanov, Congreve, Cowen-Rivers, Cowie, Figurnov, Fuchs, Gladman, Jain, Khan, Low, Perlin, Potapenko, Savy, Singh, Stecula, Thillaisundaram, Tong, Yakneen, Zhong, Zielinski, {\v Z}{\'\i}dek, Bapst, Kohli, Jaderberg, Hassabis, and Jumper]{af3}
Abramson, J., Adler, J., Dunger, J., Evans, R., Green, T., Pritzel, A., Ronneberger, O., Willmore, L., Ballard, A.~J., Bambrick, J., Bodenstein, S.~W., Evans, D.~A., Hung, C.-C., O'Neill, M., Reiman, D., Tunyasuvunakool, K., Wu, Z., {\v Z}emgulyt{\.e}, A., Arvaniti, E., Beattie, C., Bertolli, O., Bridgland, A., Cherepanov, A., Congreve, M., Cowen-Rivers, A.~I., Cowie, A., Figurnov, M., Fuchs, F.~B., Gladman, H., Jain, R., Khan, Y.~A., Low, C. M.~R., Perlin, K., Potapenko, A., Savy, P., Singh, S., Stecula, A., Thillaisundaram, A., Tong, C., Yakneen, S., Zhong, E.~D., Zielinski, M., {\v Z}{\'\i}dek, A., Bapst, V., Kohli, P., Jaderberg, M., Hassabis, D., and Jumper, J.~M.
\newblock Accurate structure prediction of biomolecular interactions with alphafold 3.
\newblock \emph{Nature}, 630\penalty0 (8016):\penalty0 493--500, 2024.
\newblock \doi{10.1038/s41586-024-07487-w}.
\newblock URL \url{https://doi.org/10.1038/s41586-024-07487-w}.

\bibitem[Ahdritz et~al.(2024)Ahdritz, Bouatta, Floristean, Kadyan, Xia, Gerecke, O’Donnell, Berenberg, Fisk, Zanichelli, et~al.]{openfold}
Ahdritz, G., Bouatta, N., Floristean, C., Kadyan, S., Xia, Q., Gerecke, W., O’Donnell, T.~J., Berenberg, D., Fisk, I., Zanichelli, N., et~al.
\newblock Openfold: Retraining alphafold2 yields new insights into its learning mechanisms and capacity for generalization.
\newblock \emph{Nature Methods}, 21\penalty0 (8):\penalty0 1514--1524, 2024.

\bibitem[Akbar et~al.(2022)Akbar, Bashour, Rawat, Robert, Smorodina, Cotet, Flem-Karlsen, Frank, Mehta, Vu, et~al.]{akbar2022progress}
Akbar, R., Bashour, H., Rawat, P., Robert, P.~A., Smorodina, E., Cotet, T.-S., Flem-Karlsen, K., Frank, R., Mehta, B.~B., Vu, M.~H., et~al.
\newblock Progress and challenges for the machine learning-based design of fit-for-purpose monoclonal antibodies.
\newblock In \emph{MAbs}, volume~14, pp.\  2008790. Taylor \& Francis, 2022.

\bibitem[Al-Lazikani et~al.(1997)Al-Lazikani, Lesk, and Chothia]{al1997standard}
Al-Lazikani, B., Lesk, A.~M., and Chothia, C.
\newblock Standard conformations for the canonical structures of immunoglobulins.
\newblock \emph{Journal of molecular biology}, 273\penalty0 (4):\penalty0 927--948, 1997.

\bibitem[Alford et~al.(2017)Alford, Leaver-Fay, Jeliazkov, O’Meara, DiMaio, Park, Shapovalov, Renfrew, Mulligan, Kappel, et~al.]{alford2017rosetta}
Alford, R.~F., Leaver-Fay, A., Jeliazkov, J.~R., O’Meara, M.~J., DiMaio, F.~P., Park, H., Shapovalov, M.~V., Renfrew, P.~D., Mulligan, V.~K., Kappel, K., et~al.
\newblock The rosetta all-atom energy function for macromolecular modeling and design.
\newblock \emph{Journal of chemical theory and computation}, 13\penalty0 (6):\penalty0 3031--3048, 2017.

\bibitem[Baek et~al.(2021)Baek, DiMaio, Anishchenko, Dauparas, Ovchinnikov, Lee, Wang, Cong, Kinch, Schaeffer, et~al.]{baek2021accurate}
Baek, M., DiMaio, F., Anishchenko, I., Dauparas, J., Ovchinnikov, S., Lee, G.~R., Wang, J., Cong, Q., Kinch, L.~N., Schaeffer, R.~D., et~al.
\newblock Accurate prediction of protein structures and interactions using a three-track neural network.
\newblock \emph{Science}, 373\penalty0 (6557):\penalty0 871--876, 2021.

\bibitem[Batool et~al.(2019)Batool, Ahmad, and Choi]{batool2019structure}
Batool, M., Ahmad, B., and Choi, S.
\newblock A structure-based drug discovery paradigm.
\newblock \emph{International journal of molecular sciences}, 20\penalty0 (11):\penalty0 2783, 2019.

\bibitem[Bengio et~al.(2009)Bengio, Louradour, Collobert, and Weston]{curriculum}
Bengio, Y., Louradour, J., Collobert, R., and Weston, J.
\newblock Curriculum learning.
\newblock In \emph{Proceedings of the 26th Annual International Conference on Machine Learning}, ICML '09, pp.\  41–48, New York, NY, USA, 2009. Association for Computing Machinery.
\newblock ISBN 9781605585161.
\newblock \doi{10.1145/1553374.1553380}.
\newblock URL \url{https://doi.org/10.1145/1553374.1553380}.

\bibitem[Berman et~al.(2000)Berman, Westbrook, Feng, Gilliland, Bhat, Weissig, Shindyalov, and Bourne]{pdb}
Berman, H.~M., Westbrook, J., Feng, Z., Gilliland, G., Bhat, T.~N., Weissig, H., Shindyalov, I.~N., and Bourne, P.~E.
\newblock The protein data bank.
\newblock \emph{Nucleic acids research}, 28\penalty0 (1):\penalty0 235--242, 2000.

\bibitem[Bernardi et~al.(2015)Bernardi, Melo, and Schulten]{bernardi2015enhanced}
Bernardi, R.~C., Melo, M.~C., and Schulten, K.
\newblock Enhanced sampling techniques in molecular dynamics simulations of biological systems.
\newblock \emph{Biochimica et Biophysica Acta (BBA)-General Subjects}, 1850\penalty0 (5):\penalty0 872--877, 2015.

\bibitem[Borkakoti \& Thornton(2023)Borkakoti and Thornton]{borkakoti2023alphafold2}
Borkakoti, N. and Thornton, J.~M.
\newblock Alphafold2 protein structure prediction: Implications for drug discovery.
\newblock \emph{Current opinion in structural biology}, 78:\penalty0 102526, 2023.

\bibitem[Borkakoti et~al.(2025)Borkakoti, Ribeiro, and Thornton]{borkakoti2025structural}
Borkakoti, N., Ribeiro, A.~J., and Thornton, J.~M.
\newblock A structural perspective on enzymes and their catalytic mechanisms.
\newblock \emph{Current Opinion in Structural Biology}, 92:\penalty0 103040, 2025.

\bibitem[Chakravarty \& Porter(2022)Chakravarty and Porter]{chakravarty2022alphafold2}
Chakravarty, D. and Porter, L.~L.
\newblock Alphafold2 fails to predict protein fold switching.
\newblock \emph{Protein Science}, 31\penalty0 (6):\penalty0 e4353, 2022.

\bibitem[Cheng et~al.(2014)Cheng, Schaeffer, Liao, Kinch, Pei, Shi, Kim, and Grishin]{cheng2014ecod}
Cheng, H., Schaeffer, R.~D., Liao, Y., Kinch, L.~N., Pei, J., Shi, S., Kim, B.-H., and Grishin, N.~V.
\newblock Ecod: an evolutionary classification of protein domains.
\newblock \emph{PLoS computational biology}, 10\penalty0 (12):\penalty0 e1003926, 2014.

\bibitem[Di~Noia \& Neuberger(2007)Di~Noia and Neuberger]{di2007molecular}
Di~Noia, J.~M. and Neuberger, M.~S.
\newblock Molecular mechanisms of antibody somatic hypermutation.
\newblock \emph{Annu. Rev. Biochem.}, 76\penalty0 (1):\penalty0 1--22, 2007.

\bibitem[Dissanayake et~al.(2025)Dissanayake, Roy, Maghsoud, Polara, Debnath, and Cisneros]{dissanayake2025computational}
Dissanayake, U.~C., Roy, A., Maghsoud, Y., Polara, S., Debnath, T., and Cisneros, G.~A.
\newblock Computational studies on the functional and structural impact of pathogenic mutations in enzymes.
\newblock \emph{Protein Science}, 34\penalty0 (4):\penalty0 e70081, 2025.

\bibitem[Dreyer(2025)]{zenodo}
Dreyer, F.~A.
\newblock Ibex: Conformation-aware structure prediction of antigen-recognizing immune proteins, July 2025.
\newblock URL \url{https://doi.org/10.5281/zenodo.15866555}.

\bibitem[Dunbar \& Deane(2016)Dunbar and Deane]{anarci}
Dunbar, J. and Deane, C.~M.
\newblock Anarci: antigen receptor numbering and receptor classification.
\newblock \emph{Bioinformatics}, 32\penalty0 (2):\penalty0 298--300, 2016.

\bibitem[Dunbar et~al.(2013)Dunbar, Fuchs, Shi, and Deane]{abangle}
Dunbar, J., Fuchs, A., Shi, J., and Deane, C.~M.
\newblock Abangle: characterising the vh--vl orientation in antibodies.
\newblock \emph{Protein Engineering, Design \& Selection}, 26\penalty0 (10):\penalty0 611--620, 2013.

\bibitem[Dunbar et~al.(2014)Dunbar, Krawczyk, Leem, Baker, Fuchs, Georges, Shi, and Deane]{sabdab}
Dunbar, J., Krawczyk, K., Leem, J., Baker, T., Fuchs, A., Georges, G., Shi, J., and Deane, C.~M.
\newblock Sabdab: the structural antibody database.
\newblock \emph{Nucleic acids research}, 42\penalty0 (D1):\penalty0 D1140--D1146, 2014.

\bibitem[{ESM Team}(2024)]{esmc}
{ESM Team}.
\newblock Esm cambrian: Revealing the mysteries of proteins with unsupervised learning, 2024.
\newblock URL \url{https://evolutionaryscale.ai/blog/esm-cambrian}.

\bibitem[Evans et~al.(2021)Evans, O’Neill, Pritzel, Antropova, Senior, Green, {\v{Z}}{\'\i}dek, Bates, Blackwell, Yim, et~al.]{af2m}
Evans, R., O’Neill, M., Pritzel, A., Antropova, N., Senior, A., Green, T., {\v{Z}}{\'\i}dek, A., Bates, R., Blackwell, S., Yim, J., et~al.
\newblock Protein complex prediction with alphafold-multimer.
\newblock \emph{biorxiv}, pp.\  2021--10, 2021.

\bibitem[Garces et~al.(2020)Garces, Mohr, Zhang, Huang, Chen, King, Xu, and Wang]{garces2020molecular}
Garces, F., Mohr, C., Zhang, L., Huang, C.-S., Chen, Q., King, C., Xu, C., and Wang, Z.
\newblock Molecular insight into recognition of the cgrpr complex by migraine prevention therapy aimovig (erenumab).
\newblock \emph{Cell Reports}, 30\penalty0 (6):\penalty0 1714--1723, 2020.

\bibitem[Greenshields-Watson et~al.(2025)Greenshields-Watson, Vavourakis, Spoendlin, Cagiada, and Deane]{GREENSHIELDSWATSON2025102983}
Greenshields-Watson, A., Vavourakis, O., Spoendlin, F.~C., Cagiada, M., and Deane, C.~M.
\newblock Challenges and compromises: Predicting unbound antibody structures with deep learning.
\newblock \emph{Current Opinion in Structural Biology}, 90:\penalty0 102983, 2025.
\newblock ISSN 0959-440X.
\newblock \doi{https://doi.org/10.1016/j.sbi.2025.102983}.
\newblock URL \url{https://www.sciencedirect.com/science/article/pii/S0959440X25000016}.

\bibitem[Hitawala \& Gray(2025)Hitawala and Gray]{hitawala2025does}
Hitawala, F.~N. and Gray, J.~J.
\newblock What does alphafold3 learn about antigen and nanobody docking, and what remains unsolved?
\newblock \emph{bioRxiv}, pp.\  2024--09, 2025.

\bibitem[Isert et~al.(2023)Isert, Atz, and Schneider]{isert2023structure}
Isert, C., Atz, K., and Schneider, G.
\newblock Structure-based drug design with geometric deep learning.
\newblock \emph{Current Opinion in Structural Biology}, 79:\penalty0 102548, 2023.

\bibitem[Jing et~al.(2024)Jing, Berger, and Jaakkola]{jing2024alphafold}
Jing, B., Berger, B., and Jaakkola, T.
\newblock Alphafold meets flow matching for generating protein ensembles.
\newblock In \emph{Forty-first International Conference on Machine Learning}, 2024.

\bibitem[Jumper et~al.(2021)Jumper, Evans, Pritzel, Green, Figurnov, Ronneberger, Tunyasuvunakool, Bates, {\v{Z}}{\'\i}dek, Potapenko, et~al.]{af2}
Jumper, J., Evans, R., Pritzel, A., Green, T., Figurnov, M., Ronneberger, O., Tunyasuvunakool, K., Bates, R., {\v{Z}}{\'\i}dek, A., Potapenko, A., et~al.
\newblock Highly accurate protein structure prediction with alphafold.
\newblock \emph{nature}, 596\penalty0 (7873):\penalty0 583--589, 2021.

\bibitem[Kabsch \& Sander(1983)Kabsch and Sander]{dssp}
Kabsch, W. and Sander, C.
\newblock Dictionary of protein secondary structure: Pattern recognition of hydrogen-bonded and geometrical features.
\newblock \emph{Biopolymers}, 22\penalty0 (12):\penalty0 2577--2637, 1983.
\newblock \doi{https://doi.org/10.1002/bip.360221211}.
\newblock URL \url{https://onlinelibrary.wiley.com/doi/abs/10.1002/bip.360221211}.

\bibitem[Kallenborn et~al.(2024)Kallenborn, Chacon, Hundt, Sirelkhatim, Didi, Cha, Dallago, Mirdita, Schmidt, and Steinegger]{mmseqs2gpu}
Kallenborn, F., Chacon, A., Hundt, C., Sirelkhatim, H., Didi, K., Cha, S., Dallago, C., Mirdita, M., Schmidt, B., and Steinegger, M.
\newblock Gpu-accelerated homology search with mmseqs2.
\newblock \emph{bioRxiv}, pp.\  2024--11, 2024.

\bibitem[Karplus \& McCammon(2002)Karplus and McCammon]{karplus2002molecular}
Karplus, M. and McCammon, J.~A.
\newblock Molecular dynamics simulations of biomolecules.
\newblock \emph{Nature structural biology}, 9\penalty0 (9):\penalty0 646--652, 2002.

\bibitem[Kenlay et~al.(2024)Kenlay, Dreyer, Cutting, Nissley, and Deane]{abb3}
Kenlay, H., Dreyer, F.~A., Cutting, D., Nissley, D., and Deane, C.~M.
\newblock Abodybuilder3: improved and scalable antibody structure predictions.
\newblock \emph{Bioinformatics}, 40\penalty0 (10):\penalty0 btae576, 10 2024.
\newblock ISSN 1367-4811.
\newblock \doi{10.1093/bioinformatics/btae576}.
\newblock URL \url{https://doi.org/10.1093/bioinformatics/btae576}.

\bibitem[Kovaltsuk et~al.(2018)Kovaltsuk, Leem, Kelm, Snowden, Deane, and Krawczyk]{oas1}
Kovaltsuk, A., Leem, J., Kelm, S., Snowden, J., Deane, C.~M., and Krawczyk, K.
\newblock Observed antibody space: a resource for data mining next-generation sequencing of antibody repertoires.
\newblock \emph{The Journal of Immunology}, 201\penalty0 (8):\penalty0 2502--2509, 2018.

\bibitem[Kovtun et~al.(2024)Kovtun, Akdel, Goncearenco, Zhou, Holt, Baugher, Lin, Adeshina, Castiglione, Wang, et~al.]{kovtun2024pinder}
Kovtun, D., Akdel, M., Goncearenco, A., Zhou, G., Holt, G., Baugher, D., Lin, D., Adeshina, Y., Castiglione, T., Wang, X., et~al.
\newblock Pinder: The protein interaction dataset and evaluation resource.
\newblock \emph{bioRxiv}, pp.\  2024--07, 2024.

\bibitem[Kreitz et~al.(2023)Kreitz, Friedrich, Guru, Lash, Saito, Macrae, and Zhang]{kreitz2023programmable}
Kreitz, J., Friedrich, M.~J., Guru, A., Lash, B., Saito, M., Macrae, R.~K., and Zhang, F.
\newblock Programmable protein delivery with a bacterial contractile injection system.
\newblock \emph{Nature}, 616\penalty0 (7956):\penalty0 357--364, 2023.

\bibitem[Lee et~al.(2022)Lee, Yadollahpour, Watkins, Frey, Leaver-Fay, Ra, Cho, Gligorijevi{\'c}, Regev, and Bonneau]{lee2022equifold}
Lee, J.~H., Yadollahpour, P., Watkins, A., Frey, N.~C., Leaver-Fay, A., Ra, S., Cho, K., Gligorijevi{\'c}, V., Regev, A., and Bonneau, R.
\newblock Equifold: Protein structure prediction with a novel coarse-grained structure representation.
\newblock \emph{Biorxiv}, pp.\  2022--10, 2022.

\bibitem[Leem et~al.(2018)Leem, de~Oliveira, Krawczyk, and Deane]{stcrdab}
Leem, J., de~Oliveira, S. H.~P., Krawczyk, K., and Deane, C.~M.
\newblock Stcrdab: the structural t-cell receptor database.
\newblock \emph{Nucleic acids research}, 46\penalty0 (D1):\penalty0 D406--D412, 2018.

\bibitem[Lewis et~al.(2024)Lewis, Hempel, Jim{\'e}nez-Luna, Gastegger, Xie, Foong, Satorras, Abdin, Veeling, Zaporozhets, Chen, Yang, Schneuing, Nigam, Barbero, Stimper, Campbell, Yim, Lienen, Shi, Zheng, Schulz, Munir, Clementi, and No{\'e}]{BioEmu2024}
Lewis, S., Hempel, T., Jim{\'e}nez-Luna, J., Gastegger, M., Xie, Y., Foong, A. Y.~K., Satorras, V.~G., Abdin, O., Veeling, B.~S., Zaporozhets, I., Chen, Y., Yang, S., Schneuing, A., Nigam, J., Barbero, F., Stimper, V., Campbell, A., Yim, J., Lienen, M., Shi, Y., Zheng, S., Schulz, H., Munir, U., Clementi, C., and No{\'e}, F.
\newblock Scalable emulation of protein equilibrium ensembles with generative deep learning.
\newblock \emph{bioRxiv}, 2024.
\newblock \doi{10.1101/2024.12.05.626885}.

\bibitem[Lin et~al.(2024)Lin, Luo, Liu, and Jin]{lin2024comprehensive}
Lin, B., Luo, X., Liu, Y., and Jin, X.
\newblock A comprehensive review and comparison of existing computational methods for protein function prediction.
\newblock \emph{Briefings in Bioinformatics}, 25\penalty0 (4):\penalty0 bbae289, 2024.

\bibitem[Lin et~al.(2023)Lin, Akin, Rao, Hie, Zhu, Lu, Smetanin, Verkuil, Kabeli, Shmueli, et~al.]{esmfold}
Lin, Z., Akin, H., Rao, R., Hie, B., Zhu, Z., Lu, W., Smetanin, N., Verkuil, R., Kabeli, O., Shmueli, Y., et~al.
\newblock Evolutionary-scale prediction of atomic-level protein structure with a language model.
\newblock \emph{Science}, 379\penalty0 (6637):\penalty0 1123--1130, 2023.

\bibitem[Loshchilov \& Hutter(2017)Loshchilov and Hutter]{adamW}
Loshchilov, I. and Hutter, F.
\newblock Decoupled weight decay regularization.
\newblock \emph{arXiv preprint arXiv:1711.05101}, 2017.

\bibitem[Loukas et~al.(2024)Loukas, Martinkus, Wagstaff, and Cho]{loukas2024generalizing}
Loukas, A., Martinkus, K., Wagstaff, E., and Cho, K.
\newblock Generalizing to any diverse distribution: uniformity, gentle finetuning and rebalancing.
\newblock \emph{arXiv preprint arXiv:2410.05980}, 2024.

\bibitem[Lu et~al.(2020)Lu, Hwang, Liu, Lee, Tsai, Li, and Wu]{lu2020development}
Lu, R.-M., Hwang, Y.-C., Liu, I.-J., Lee, C.-C., Tsai, H.-Z., Li, H.-J., and Wu, H.-C.
\newblock Development of therapeutic antibodies for the treatment of diseases.
\newblock \emph{Journal of biomedical science}, 27:\penalty0 1--30, 2020.

\bibitem[Marks et~al.(2018)Marks, Shi, and Deane]{marks2018predicting}
Marks, C., Shi, J., and Deane, C.~M.
\newblock Predicting loop conformational ensembles.
\newblock \emph{Bioinformatics}, 34\penalty0 (6):\penalty0 949--956, 2018.

\bibitem[Marks et~al.(2012)Marks, Hopf, and Sander]{marks2012protein}
Marks, D.~S., Hopf, T.~A., and Sander, C.
\newblock Protein structure prediction from sequence variation.
\newblock \emph{Nature biotechnology}, 30\penalty0 (11):\penalty0 1072--1080, 2012.

\bibitem[Narciso et~al.(2011)Narciso, Uy, Cabang, Chavez, Pablo, Padilla-Concepcion, and Padlan]{NARCISO2011435}
Narciso, J. E.~T., Uy, I. D.~C., Cabang, A.~B., Chavez, J. F.~C., Pablo, J. L.~B., Padilla-Concepcion, G.~P., and Padlan, E.~A.
\newblock Analysis of the antibody structure based on high-resolution crystallographic studies.
\newblock \emph{New Biotechnology}, 28\penalty0 (5):\penalty0 435--447, 2011.
\newblock ISSN 1871-6784.
\newblock \doi{https://doi.org/10.1016/j.nbt.2011.03.012}.
\newblock URL \url{https://www.sciencedirect.com/science/article/pii/S1871678411000744}.
\newblock Antibodies: From Basics to Therapeutics.

\bibitem[North et~al.(2011)North, Lehmann, and Dunbrack~Jr]{north}
North, B., Lehmann, A., and Dunbrack~Jr, R.~L.
\newblock A new clustering of antibody cdr loop conformations.
\newblock \emph{Journal of molecular biology}, 406\penalty0 (2):\penalty0 228--256, 2011.

\bibitem[Olsen et~al.(2022)Olsen, Boyles, and Deane]{oas2}
Olsen, T.~H., Boyles, F., and Deane, C.~M.
\newblock Observed antibody space: A diverse database of cleaned, annotated, and translated unpaired and paired antibody sequences.
\newblock \emph{Protein Science}, 31\penalty0 (1):\penalty0 141--146, 2022.

\bibitem[Passaro et~al.(2025)Passaro, Corso, Wohlwend, Reveiz, Thaler, Ram~Somnath, Getz, Portnoi, Roy, Stark, et~al.]{boltz2}
Passaro, S., Corso, G., Wohlwend, J., Reveiz, M., Thaler, S., Ram~Somnath, V., Getz, N., Portnoi, T., Roy, J., Stark, H., et~al.
\newblock Boltz-2: Towards accurate and efficient binding affinity prediction.
\newblock \emph{bioRxiv}, pp.\  2025--06, 2025.

\bibitem[Quast et~al.(2025)Quast, Abanades, Guloglu, Karuppiah, Harper, Raybould, and Deane]{tcrbuilder2p}
Quast, N.~P., Abanades, B., Guloglu, B., Karuppiah, V., Harper, S., Raybould, M.~I., and Deane, C.~M.
\newblock T-cell receptor structures and predictive models reveal comparable alpha and beta chain structural diversity despite differing genetic complexity.
\newblock \emph{Communications Biology}, 8\penalty0 (1):\penalty0 362, 2025.

\bibitem[Rini et~al.(1992)Rini, Schulze-Gahmen, and Wilson]{rini1992structural}
Rini, J.~M., Schulze-Gahmen, U., and Wilson, I.~A.
\newblock Structural evidence for induced fit as a mechanism for antibody-antigen recognition.
\newblock \emph{Science}, 255\penalty0 (5047):\penalty0 959--965, 1992.

\bibitem[Ruffolo et~al.(2023)Ruffolo, Chu, Mahajan, and Gray]{ruffolo2023fast}
Ruffolo, J.~A., Chu, L.-S., Mahajan, S.~P., and Gray, J.~J.
\newblock Fast, accurate antibody structure prediction from deep learning on massive set of natural antibodies.
\newblock \emph{Nature communications}, 14\penalty0 (1):\penalty0 2389, 2023.

\bibitem[Schafer \& Porter(2025)Schafer and Porter]{schafer2025alphafold2}
Schafer, J.~W. and Porter, L.~L.
\newblock Alphafold2's training set powers its predictions of some fold-switched conformations.
\newblock \emph{Protein Science}, 34\penalty0 (4):\penalty0 e70105, 2025.

\bibitem[Spoendlin et~al.(2025)Spoendlin, Fern{\'a}ndez-Quintero, Raghavan, Turner, Gharpure, Loeffler, Wong, Bujotzek, Georges, Ward, et~al.]{spoendlin2025predicting}
Spoendlin, F.~C., Fern{\'a}ndez-Quintero, M.~L., Raghavan, S.~S., Turner, H.~L., Gharpure, A., Loeffler, J.~R., Wong, W.~K., Bujotzek, A., Georges, G., Ward, A.~B., et~al.
\newblock Predicting the conformational flexibility of antibody and t-cell receptor cdrs.
\newblock \emph{bioRxiv}, pp.\  2025--03, 2025.

\bibitem[Steinegger \& S{\"o}ding(2017)Steinegger and S{\"o}ding]{mmseqs2}
Steinegger, M. and S{\"o}ding, J.
\newblock Mmseqs2 enables sensitive protein sequence searching for the analysis of massive data sets.
\newblock \emph{Nature biotechnology}, 35\penalty0 (11):\penalty0 1026--1028, 2017.

\bibitem[team et~al.(2024)team, Boitreaud, Dent, McPartlon, Meier, Reis, Rogozhonikov, and Wu]{chai}
team, C.~D., Boitreaud, J., Dent, J., McPartlon, M., Meier, J., Reis, V., Rogozhonikov, A., and Wu, K.
\newblock Chai-1: Decoding the molecular interactions of life.
\newblock \emph{bioRxiv}, 2024.
\newblock \doi{10.1101/2024.10.10.615955}.
\newblock URL \url{https://www.biorxiv.org/content/early/2024/10/11/2024.10.10.615955}.

\bibitem[Tsuchiya \& Mizuguchi(2016)Tsuchiya and Mizuguchi]{tsuchiya2016diversity}
Tsuchiya, Y. and Mizuguchi, K.
\newblock The diversity of h 3 loops determines the antigen-binding tendencies of antibody cdr loops.
\newblock \emph{Protein Science}, 25\penalty0 (4):\penalty0 815--825, 2016.

\bibitem[Wohlwend et~al.(2024)Wohlwend, Corso, Passaro, Reveiz, Leidal, Swiderski, Portnoi, Chinn, Silterra, Jaakkola, and Barzilay]{boltz}
Wohlwend, J., Corso, G., Passaro, S., Reveiz, M., Leidal, K., Swiderski, W., Portnoi, T., Chinn, I., Silterra, J., Jaakkola, T., and Barzilay, R.
\newblock Boltz-1 democratizing biomolecular interaction modeling.
\newblock \emph{bioRxiv}, 2024.
\newblock \doi{10.1101/2024.11.19.624167}.
\newblock URL \url{https://www.biorxiv.org/content/early/2024/11/20/2024.11.19.624167}.

\bibitem[Xu et~al.(2025)Xu, Feng, Qiao, Wu, Shen, Cheng, Zheng, and Sun]{xu2025foldbench}
Xu, S., Feng, Q., Qiao, L., Wu, H., Shen, T., Cheng, Y., Zheng, S., and Sun, S.
\newblock Foldbench: An all-atom benchmark for biomolecular structure prediction.
\newblock \emph{bioRxiv}, pp.\  2025--05, 2025.

\end{thebibliography}
\bibliographystyle{icml2025}

\newpage
\appendix
\onecolumn
\section{Hydrogen bond contact recovery in CDR residues}
\label{app:hydrogen}
We report the recall and precision of Ibex-predicted hydrogen-bond networks within the CDRs relative to the ground truth structures. All coordinates were identically and independently relaxed 5 times using Rosetta to minimize spurious signal between experimental and predicted structures. hydrogen bonds were identified using the Rosetta ScoreFunction as those possessing $>10\%$ of the maximum hydrogen-bond energy. For each pair of experimental/predicted structures, we computed the set of unique interacting residues connected by at least one hydrogen bond, sorted by residue order to avoid double counting. The intersection of the two sets yielded true positives (TP) while false positives (FP) and false negatives (FN) were determined from their differences. Precision and recall of the ground truth hydrogen-bond interactions were computed via their standard definitions
\begin{equation}
    \mathrm{precision} = \frac{TP}{TP+FP}\,,\quad
    \mathrm{recall}=\frac{TP}{TP+FN}\,.
\end{equation}

\begin{figure}[ht]
    \centering
    \includegraphics[width=0.4\linewidth]{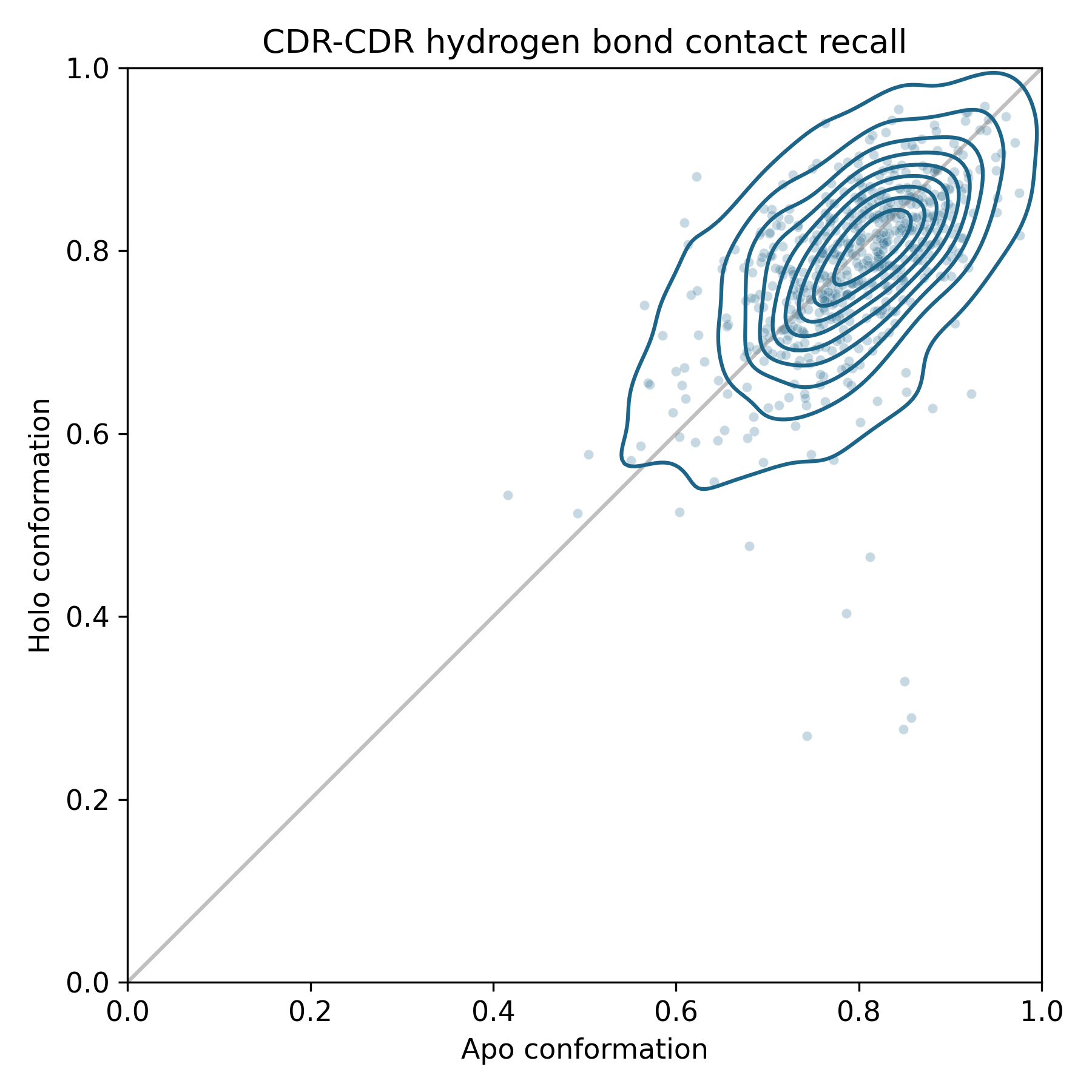}\quad
    \includegraphics[width=0.4\linewidth]{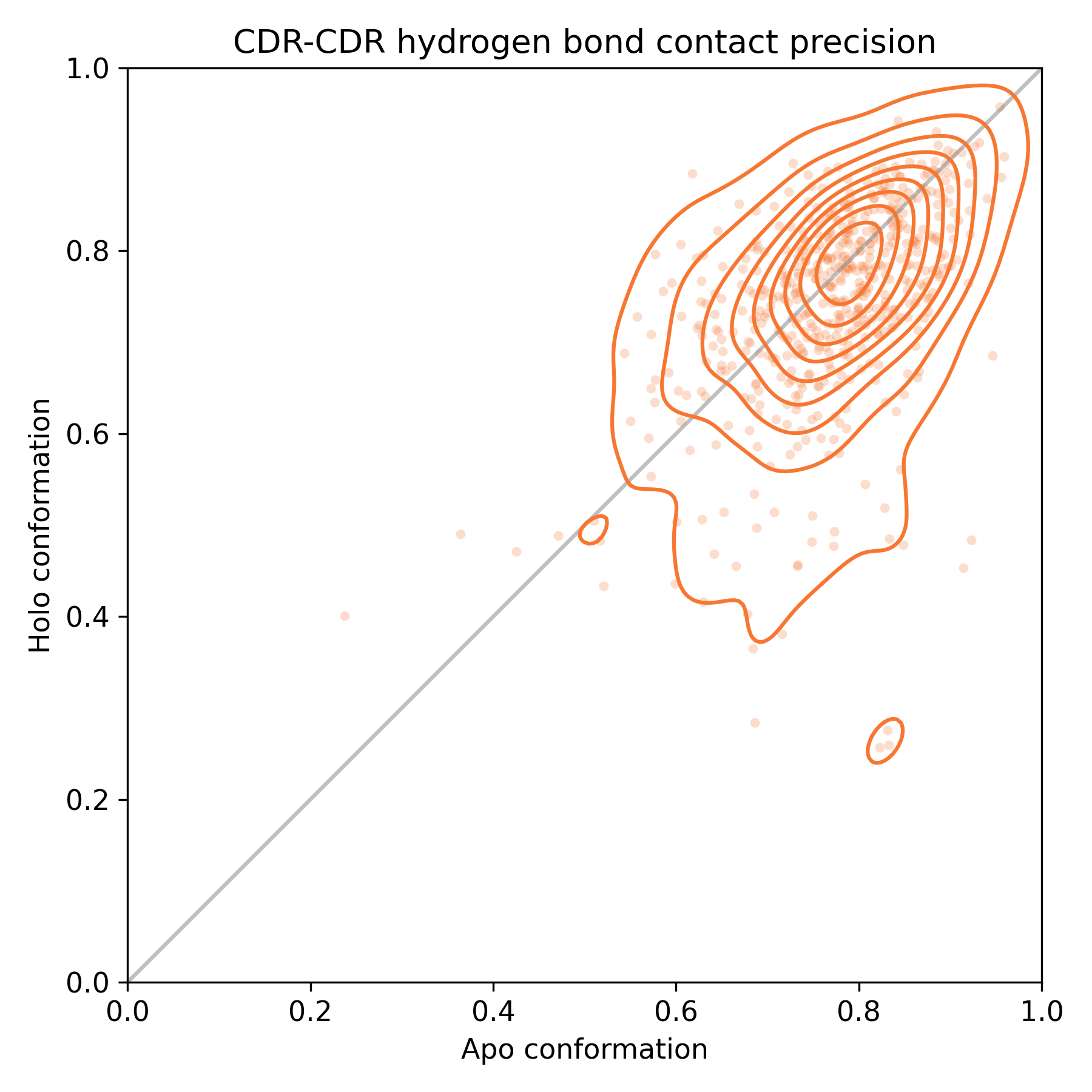}
    \caption{Recall (left) and precision (right) of hydrogen bond contacts predicted by Ibex, for pairs of \textit{apo}/\textit{holo} structures.}
    \label{fig:hydrogen-recovery}
\end{figure}

\clearpage

\section{Detailed comparison of Ibex and ImmuneBuilder}
\label{app:comparisons}
\begin{figure}[ht]
    \centering
    \includegraphics[width=1.0\linewidth]{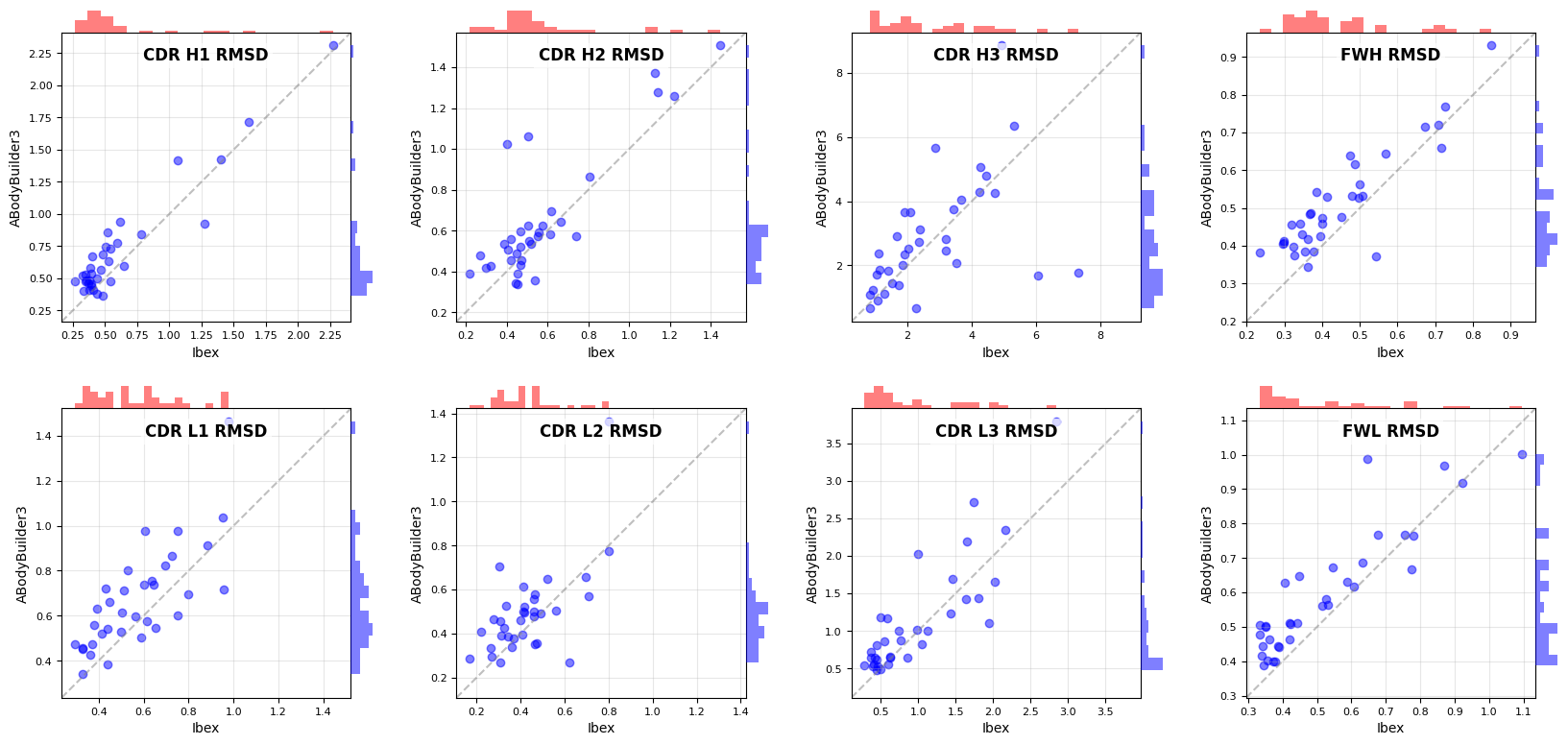}
    \caption{CDR and framework RMSD in angstrom, comparing Ibex against ABodyBuilder3 on the ImmuneBuilder test set of 34 antibodies.}
    \label{fig:comparison_fv}
\end{figure}

\begin{figure}[ht]
    \centering
    \includegraphics[width=1.0\linewidth]{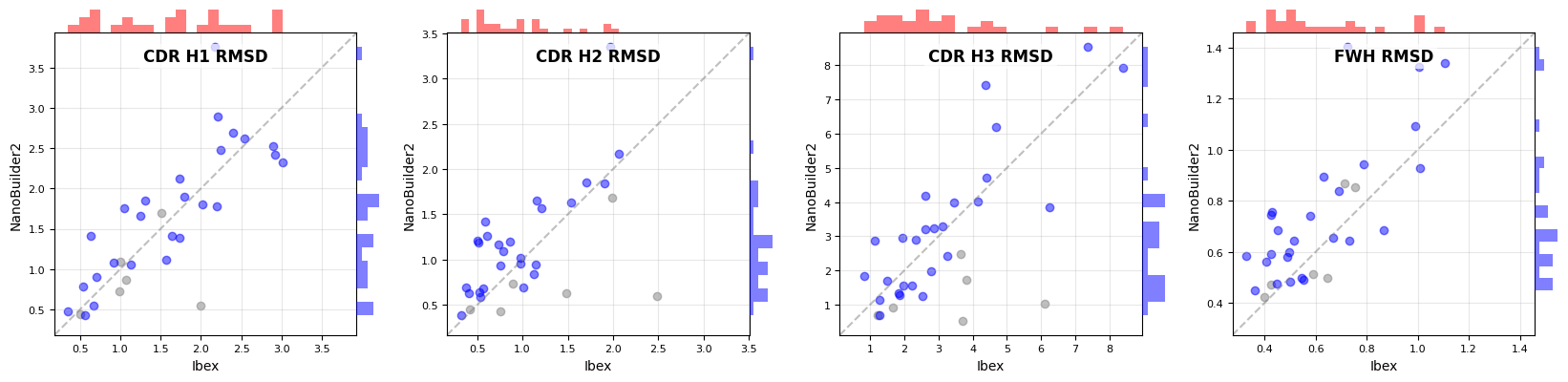}
    \caption{CDR and framework RMSD in angstrom, comparing Ibex against NanoBuilder2 on the ImmuneBuilder test set of 32 nanobodies. The 6 datapoints for which identical CDR H3 sequences were identified in the train or validation split are shown in grey and are excluded from the displayed histogram.}
    \label{fig:comparison_vhh}
\end{figure}

\begin{figure}[ht]
    \centering
    \includegraphics[width=1.0\linewidth]{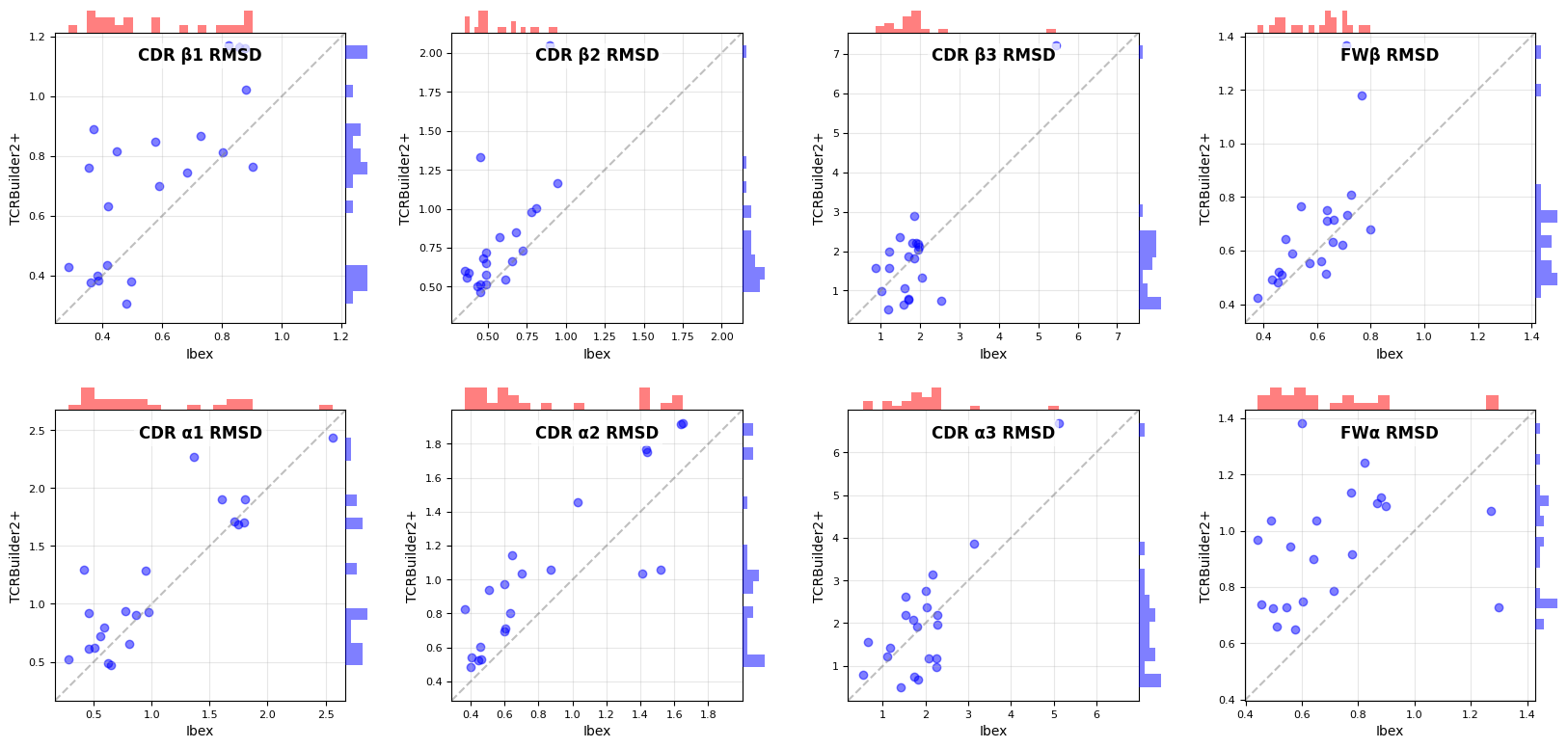}
    \caption{CDR and framework RMSD in angstrom, comparing Ibex against ABodyBuilder3 on the ImmuneBuilder test set of 21 TCRs.}
    \label{fig:comparison_tcr}
\end{figure}

We provide here a detailed view of the summary Table~\ref{tab:results}. 
To compute the RMSD, each predicted heavy and light chain is aligned independently to the corresponding ground truth chain according only to their framework residues.
In figure~\ref{fig:comparison_fv}, we show the RMSD on the ImmuneBuilder antibody test set separated by region, and comparing Ibex predictions to ABodyBuilder3. Figure~\ref{fig:comparison_vhh} gives a comparison of Ibex to NanoBuilder2 on the test set of nanobodies. Here the 6 structures for which an identical match was identified in the train or validation split of NanoBuilder2, which are PDB codes \texttt{7n4n}, \texttt{7omt}, \texttt{7q6c}, \texttt{7rg7}, \texttt{7zmv}, and \texttt{7zxu}, are shown in grey. These points were excluded from the average presented in Table~\ref{tab:results}.
In Figure~\ref{fig:comparison_tcr}, we provide a comparison of Ibex to TCRBuilder2+ on the test set of 21 TCR structures.


\clearpage

\section{Private dataset composition}
\label{app:private-dataset}
\begin{figure}[ht]
    \centering
    \includegraphics[width=0.97\linewidth]{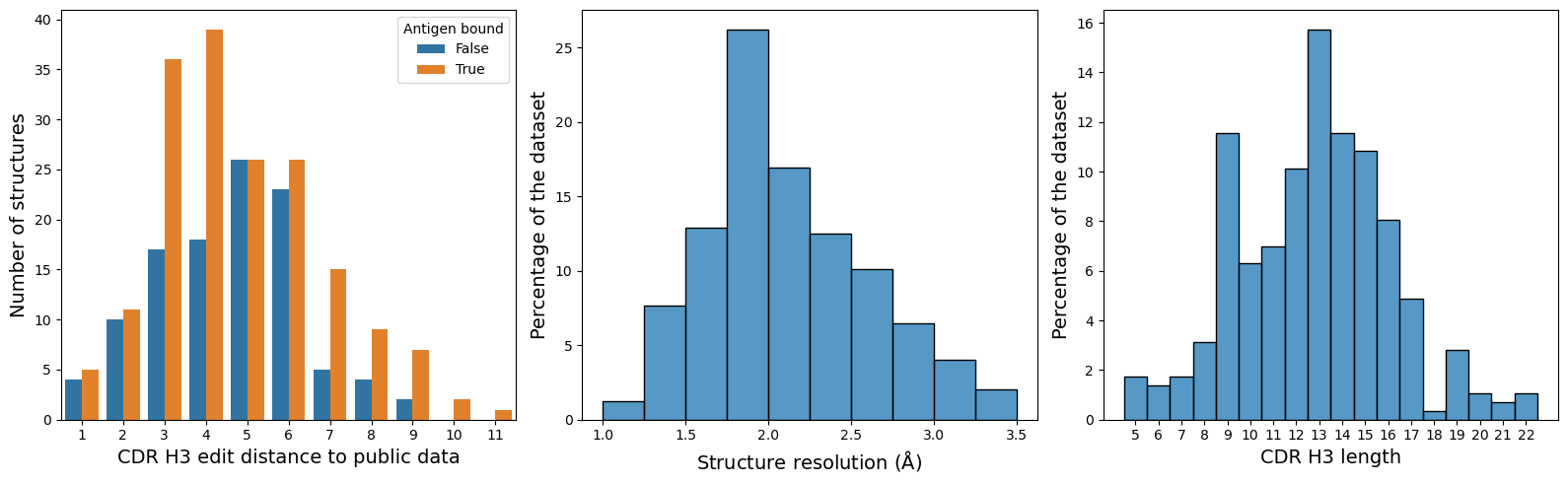}
    \caption{Distribution of CDR H3 edit distance, resolution and CDR H3 length of the private dataset used in Section~\ref{sec:private-benchmark}.}
    \label{fig:internal_data}
\end{figure}

We provide details of the private dataset of antibody structures used as benchmark in Section~\ref{sec:private-benchmark}.
The dataset comprises of 286 structures, of which 177 are bound to an antigen and 109 are in the \textit{apo} conformation.
Each structure in this dataset is at least one residue away along the CDR H3 loop sequence from any antibody in SAbDab. 
The distribution of number of structures by CDR H3 edit distance to the closest matching public datapoint, as well as the resolution and CDR H3 loop length distribution, are shown in Figure~\ref{fig:internal_data}.

\clearpage

\section{Inference compute and impact of multiple sequence alignement}
\label{app:compute}

In Table~\ref{tab:compute}, we provide the average inference time on a typical antibody variable domain for Ibex, ESMFold and Boltz-2 on an NVIDIA A10G GPU. Here Boltz-2 is evaluated with the MSAs having been precomputed and provided as an \texttt{a3m} file, and we give also the compute time used on a node of 8 NVIDIA A100 GPUs to generate these with MMseqs2-GPU. Note that inference time for MSA generation is substantially reduced when batching sequences, and we provide both batched and unbatched timings.

\begin{table}[ht]
    \centering
    \begin{tabular}{c c}
        \toprule
        Model & Average inference time\\ \midrule
        Ibex        & 0.7s on single NVIDIA A10G \\
        ESMFold & 6.7s on single NVIDIA A10G\\
        Boltz-2, no MSA  & 63.9s on single NVIDIA A10G \\
        Boltz-2, with pre-computed MSA  & 67.7s on single NVIDIA A10G \\
        MSA, per Fv on a batch of 1000 & 3.5s on $8\times$ NVIDIA A100 \\
        MSA, single Fv & 10.5min on $8\times$ NVIDIA A100  \\
        \bottomrule
    \end{tabular}
    \caption{Inference compute time for Ibex, ESMFold and Boltz-2, as well as for the calculation of MSAs with MMseqs2-GPU.}
    \label{tab:compute}
\end{table}

\begin{figure}[ht]
    \centering
    \includegraphics[width=0.5\linewidth]{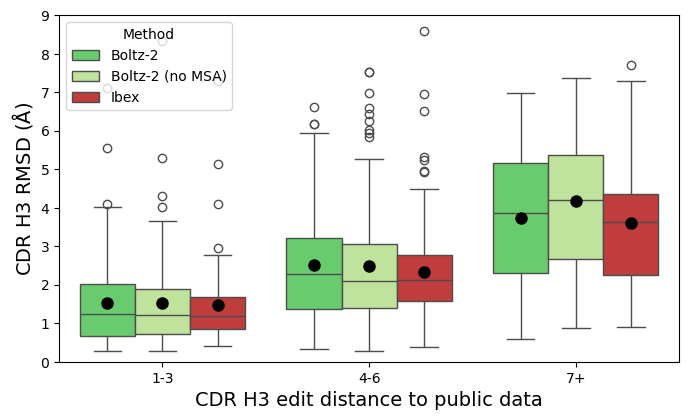}%
    \caption{CDR H3 RMSD as a function of edit distance to the closest matching H3 loop in SAbDAb, comparing Boltz-2, Boltz-2 without MSA input, and Ibex.}
    \label{fig:benchmark-nomsa}
\end{figure}

We also evaluate the performance of Boltz-2 on our private dataset when no MSA input is provided to the model. Results are shown in Figure~\ref{fig:benchmark-nomsa}, where we observe that while performance is slightly worse out-of-distribution, in the high edit distance bin, providing MSAs leads to no noticeable improvement for antibodies close to known structures.

\clearpage

\section{Ablation studies}
\label{app:ablations}

\begin{figure*}[ht]
    \centering
    \includegraphics[width=0.5\linewidth]{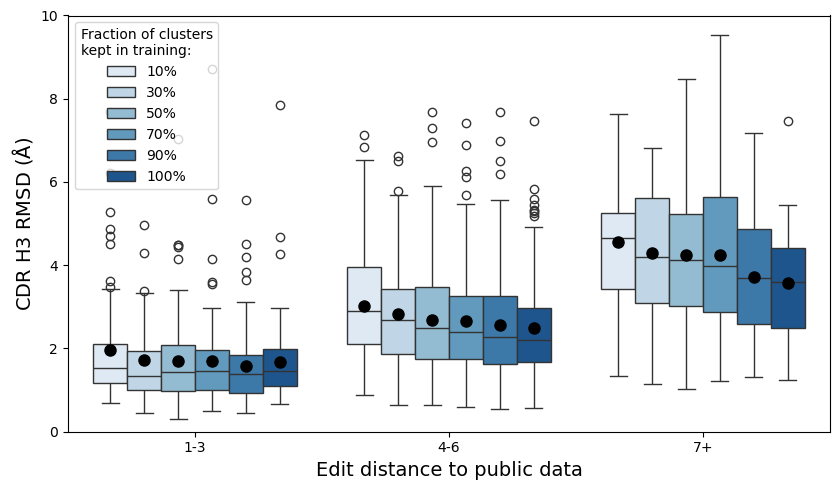}%
    \includegraphics[width=0.5\linewidth]{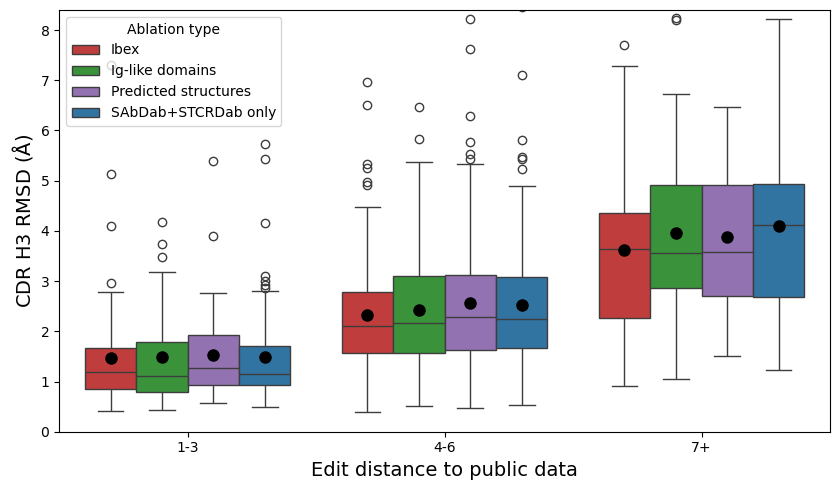}
    \caption{CDR H3 RMSD as a function of edit distance to the closest matching H3 loop in SAbDAb. Left: Impact of the training data size on model performance. We randomly subsample a fraction of training clusters and train a single checkpoint on a reduced dataset of SAbDab and STCRDab structures. Right: 
    ablation studies of the data used, training a single checkpoint on the SAbDab and STRCDab data only, as well as on a combination of the SAbDab and STCRDab data and the immunoglobulin-like data, or the predicted structures data.}
    \label{fig:ablation}
\end{figure*}

In this section we consider the impact on out-of-distribution robustness of several architecture and dataset choices.
To this end, we train several single model checkpoints for 400 epochs in each three stage.
In Figure~\ref{fig:ablation} (left), we show the impact of dataset size on model performance. Here each model is trained on only the experimental antibody and TCR datasets, SAbDab and STCRDab, randomly sub-sampling a fraction of the total available training clusters. Epoch length is rescaled for each model to correspond to the same number of iterations and training time to the model trained on 100$\%$ of the dataset. We observe a linear improvement in the high edit distance bin, suggesting that the collection of further structural data will be beneficial to model robustness and performance.
In Figure~\ref{fig:ablation} (right), we perform an ablation studies of models trained without the immunoglobulin-like data, without the predicted data, as well as trained only on SAbDab and STCRDab.
We can observe that both the predicted structures and the immunoglobulin-like data lead to improved performance at large edit distance from known public CDR H3 loops, and the full Ibex ensemble model trained on combined data sees additive improvements from both secondary datasets.

\end{document}